%% file: main.tex
\documentclass[10pt,twocolumn,letterpaper]{article}

\usepackage{cvpr}              

\input{preamble}

\definecolor{cvprblue}{rgb}{0.21,0.49,0.74}
\usepackage[pagebackref,breaklinks,colorlinks,citecolor=cvprbl ue]{hyperref}
\usepackage[accsupp]{axessibility}

\def\paperID{2632} 
\def\confName{CVPR}
\def\confYear{2024}

\title{Snap-Snap: Taking Two Images to Reconstruct \\ 3D Human Gaussians in Milliseconds}

\author{
Jia Lu$^{1}$\footnotemark[1], \; Taoran Yi$^{1}$\footnotemark[1],\; Jiemin Fang$^{2}$\footnotemark[2],\; Chen Yang$^{3}$, \; Chuiyun Wu$^{1}$, \; Wei Shen$^{3}$, \; Wenyu Liu$^1$, \\ Qi Tian$^{2}$,\; Xinggang Wang$^1$\footnotemark[2]\\
$^1$Huazhong University of Science and Technology \;\;
$^2$Huawei Inc.\\
$^3$Shanghai Jiaotong University\\
\texttt{\small\{jialu2023, taoranyi, wuchuiyun, liuwy, xgwang\}@hust.edu.cn}\\
\texttt{\small jaminfong@gmail.com}  \;\;
\texttt{\small\{ycyangchen, wei.shen\}@sjtu.edu.cn} \;\;
\texttt{\small tian.qi1@huawei.com}
}

\begin{document}
\maketitle

{
\renewcommand{\thefootnote}{\fnsymbol{footnote}}
\footnotetext[1]{Equal contribution (during internship at Huawei Inc.)}
\footnotetext[2]{Corresponding authors}
}

\begin{abstract}
Reconstructing 3D human bodies from sparse views has been an appealing topic, which is crucial to broader the related applications. In this paper, we propose a quite challenging but valuable task to reconstruct the human body from only two images, i.e., the front and back view, which can largely lower the barrier for users to create their own 3D digital humans. The main challenges lie in the difficulty of building 3D consistency and recovering missing information from the highly sparse input. We redesign a geometry reconstruction model based on foundation reconstruction models to predict consistent point clouds even input images have scarce overlaps with extensive human data training. Furthermore, an enhancement algorithm is applied to supplement the missing color information, and then the complete human point clouds with colors can be obtained, which are directly transformed into 3D Gaussians for better rendering quality. Experiments show that our method can reconstruct the entire human in 190 ms on a single NVIDIA RTX 4090, with two images at a resolution of 1024$\times$1024, demonstrating state-of-the-art performance on the THuman2.0 and cross-domain datasets. Additionally, our method can complete human reconstruction even with images captured by low-cost mobile devices, reducing the requirements for data collection. Demos and code are available at \url{https://hustvl.github.io/Snap-Snap/}.

\end{abstract}

\section{Introduction}
\label{sec: Introduction}

Human reconstruction has always been an important topic in the 3D field, which can be considered a crucial bridge between the real and digital worlds. It has broad practical application prospects, including virtual/augmented reality, games, and Metaverse. 
Professional/expensive facilities to capture human data are usually required in previous reconstruction methods~\cite{peng2021neural,zheng2024gps,yi2023generalizable,ghg_2024,li2022tava}, \emph{e.g.}, using synchronized cameras to capture the target human from many views. 
Loosening the capturing requirements allows more users to complete their own reconstruction and makes it applicable to more scenarios. 

\begin{figure}[t!]
    \centering
    \includegraphics[width=1.0\linewidth]{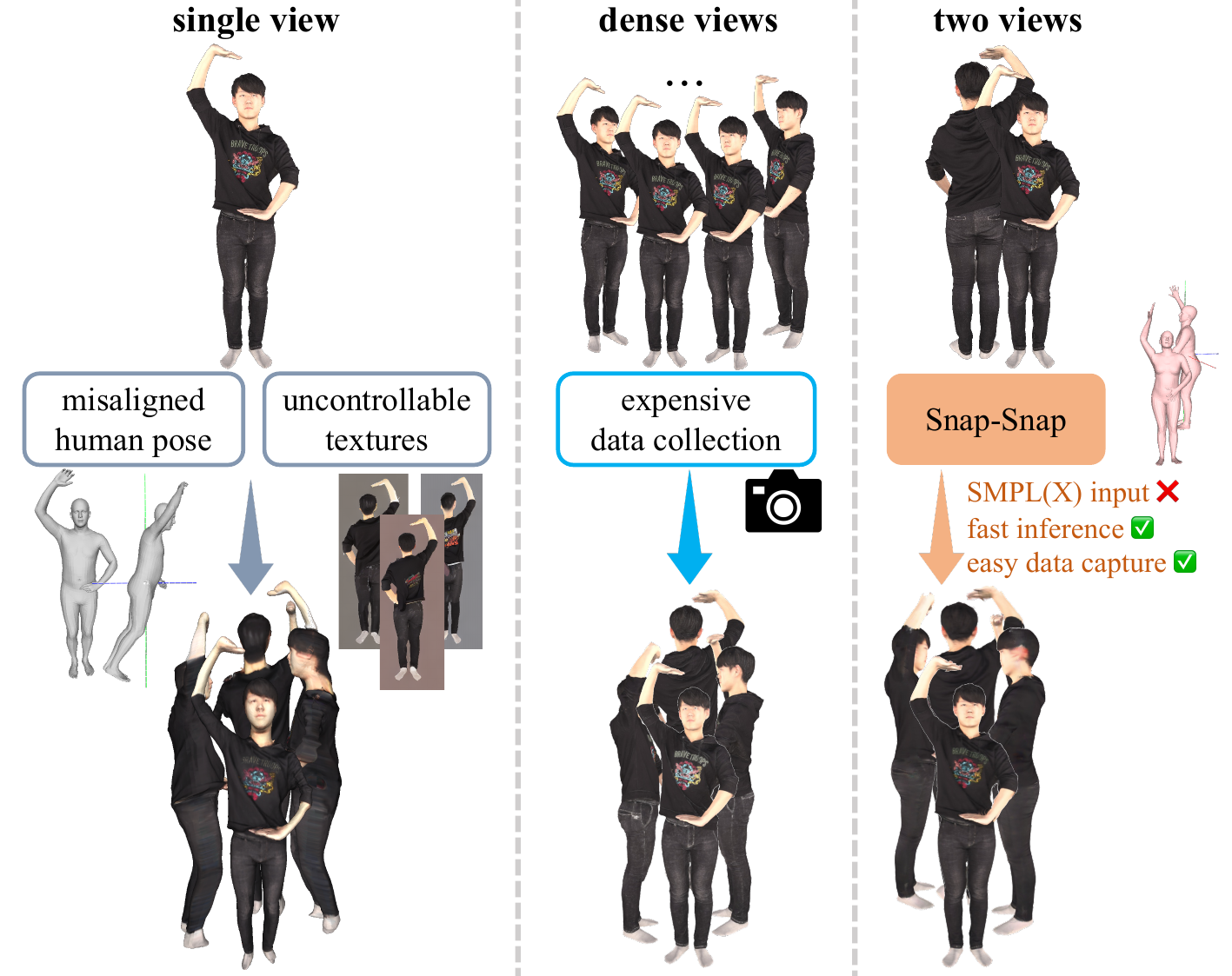}
    \caption{Under the setting of two input images, we propose a feed-forward framework named Snap-Snap which can directly predict 3D human Gaussians in milliseconds.}
    \label{fig: teaser}
\end{figure}

Reconstructing human bodies from images has always been a challenging topic and has been studied in many previous works~\cite{chen2022geometry,cheng2022generalizable,kwon2021neural,kwon2023neural,yang2024gaussianobject,shao2022floren,ghg_2024}. 
In single-view human reconstruction~\cite{ho2024sith,xue2024human3diffusion,zhang2024sifu}, human prior estimations (SMPL-X~\cite{smplx}) are frequently misaligned with real-world coordinates (\emph{e.g.}, body inclination), and generative methods~\cite{rombach2022sith_diffusion,shi2023MVDream} introduced to infer occluded regions often produce results with limited controllability. Meanwhile, human prior estimations based on limited viewpoints still suffer from inaccuracies in details, such as the hands. Although human reconstruction methods based on dense input views~\cite{zheng2024gps} achieve high-quality results, they often require expensive data acquisition setups, making them less accessible to users. Fig.~\ref{fig: teaser} provides a visual comparison of human reconstruction methods under different settings.
Under the premise of preserving a controllable and consistent human appearance, we propose to explore an extremely challenging task - reconstructing human bodies from just two images, \emph{i.e.}, the front and back view, in milliseconds. This task makes it quite easy for any user to reconstruct the desired 3D digital human, without needing professional knowledge to capture redundant images or a long wait to get the result. 
One main challenge lies in the lack of overlap between the front and back views, making it hard to establish geometry consistency for reconstruction. Additionally, the information from the two views is too limited to cover all the details of the human body. 

To tackle the above challenges, we propose to construct a fast point cloud prediction model which can reconstruct complete human point clouds from four viewpoints (front, back, left and right views) even with two input images. With training upon the human datasets, the reconstruction model based on geometry reconstruction model~\cite{dust3r_cvpr24} adapts the generalizable geometric prior to the human domain.
As the geometry reconstruction model only predicts point clouds without color information, we further design an enhancement algorithm to enhance the side-view color by wrapping.
With the above processes, a complete set of colored point clouds for the target human is obtained. To achieve better rendering quality, we transform these points into 3D Gaussians~\cite{kerbl20233d} by directly inferring corresponding Gaussian attributes. Overall, the framework is efficient enough to complete one human reconstruction at the millisecond level. All you need to do is take two human images from the front and back views, so we name our method Snap-Snap to represent the shutter sound when capturing the human. Our approach demonstrates superior reconstruction performance on several datasets.

Our contributions can be summarized as follows:
\begin{itemize}
    \item We design a feed-forward reconstruction framework, which can directly predict 3D human Gaussians from just two images in milliseconds without human prior.
    \item We redesign a geometry reconstruction model which can build human point clouds even with highly sparse input, adapting the generalizable geometric prior to the human domain. In addition, a side view enhancement algorithm is proposed to supplement the unseen information.
    \item With two-view images at a resolution of 1024$\times$1024, our method can obtain complete human reconstruction results in 190 ms and demonstrates state-of-the-art performance on the THuman2.0~\cite{yu2021function4d} and cross-domain~\cite{4d_dress, han20232k2k} datasets. The method is also shown to perform well on data acquired from low-cost mobile devices.
\end{itemize}

\section{Related Works}
\label{sec: related}
\noindent \textbf{3D Gaussian Splatting in Human.}
3D Gaussian Splatting ~\cite{kerbl20233d} explicitly reconstructs the scene with Gaussian points, achieving good reconstruction results in an efficient way. Many methods have introduced 3DGS into the field of human reconstruction, achieving state-of-the-art reconstruction results, and even high-quality animated results. 
Most of the work uses SMPL~\cite{loper2023smpl} or SMPL-X~\cite{smplx} as human prior for point clouds initialization and accepts videos as input for scene reconstruction. Most methods ~\cite{li2024gaussianbody,jena2023splatarmor,lei2024gart,kocabas2024hugs, hu2024gauhuman,hu2024gaussianavatar} accept monocular video as input for the human reconstruction, while a few methods based on multi-view video reconstruction~\cite{li2024animatable,jiang2024hifi4g,moreau2024human} have expanded in other aspects.
However, most of these works lack generalizability, requesting to be trained separately for each scene. 
In contrast, we propose a feed-forward generalizable human reconstruction method, achieving good reconstruction quality even under extremely sparse viewpoints in milliseconds.

\noindent \textbf{Generalizable Human Reconstruction.}
Generalizable Human Reconstruction aims to reconstruct the 3D human body solely through inference from input images. 
Many works have achieved generalizable reconstruction of the 3D human body under the framework of implicit representation, such as those based on pixel-aligned features ~\cite{saito2019pifu,saito2020pifuhd}, 
those based on generalizable neural voxels ~\cite{yi2023generalizable} and those based on sparse 3D keypoints ~\cite{mihajlovic2022keypointnerf}.
Meanwhile, the series of works ~\cite{xiu2022icon,xiu2023econ,huang2024tech} have generalized the extraction of human geometry solely from a single input image. HumanSplat~\cite{pan2024humansplat} infers explicit human representation from a single image with semantic cues.
Single-view human reconstruction often involves generative models, making it difficult for the reconstructed results to fully match the target subject.
Under the sparse camera setting, NHP~\cite{kwon2021neural} learns generalizable neural radiance representations with body motion prior. GHG~\cite{ghg_2024} learns generalizable human Gaussians on the 2D UV space of a human template. GPS-Gaussian~\cite{zheng2024gps} builds upon a depth estimation model, leveraging neural networks to extract Gaussian properties, and demonstrates strong results in novel view synthesis tasks.

\noindent \textbf{Points Prediction.}
With the development of neural networks~\cite{dosovitskiy2020image}, image-based scene perception capabilities have also advanced, where point clouds are extracted solely based on RGB images.
Several methods~\cite{ranftl2021vision,bian2021auto,sun2023sc} achieve scene perception based on monocular depth estimation. 
Furthermore, ~\cite{yin2021learning} utilizes point cloud neural networks to improve the estimation of point clouds.
Disparity is another representation of point cloud information in stereo vision. 
~\cite{lipson2021raft} obtains point cloud information by converting the disparity map to a depth map.
Recently, with the progress in general 3D models~\cite{weinzaepfel2023croco}, 
methods for directly predicting point clouds~\cite{dust3r_cvpr24,mast3r_arxiv24} have emerged and exhibited excellent perception capabilities.

\section{Method}

\begin{figure*}[t]
    \centering
    \includegraphics[width=1\linewidth]{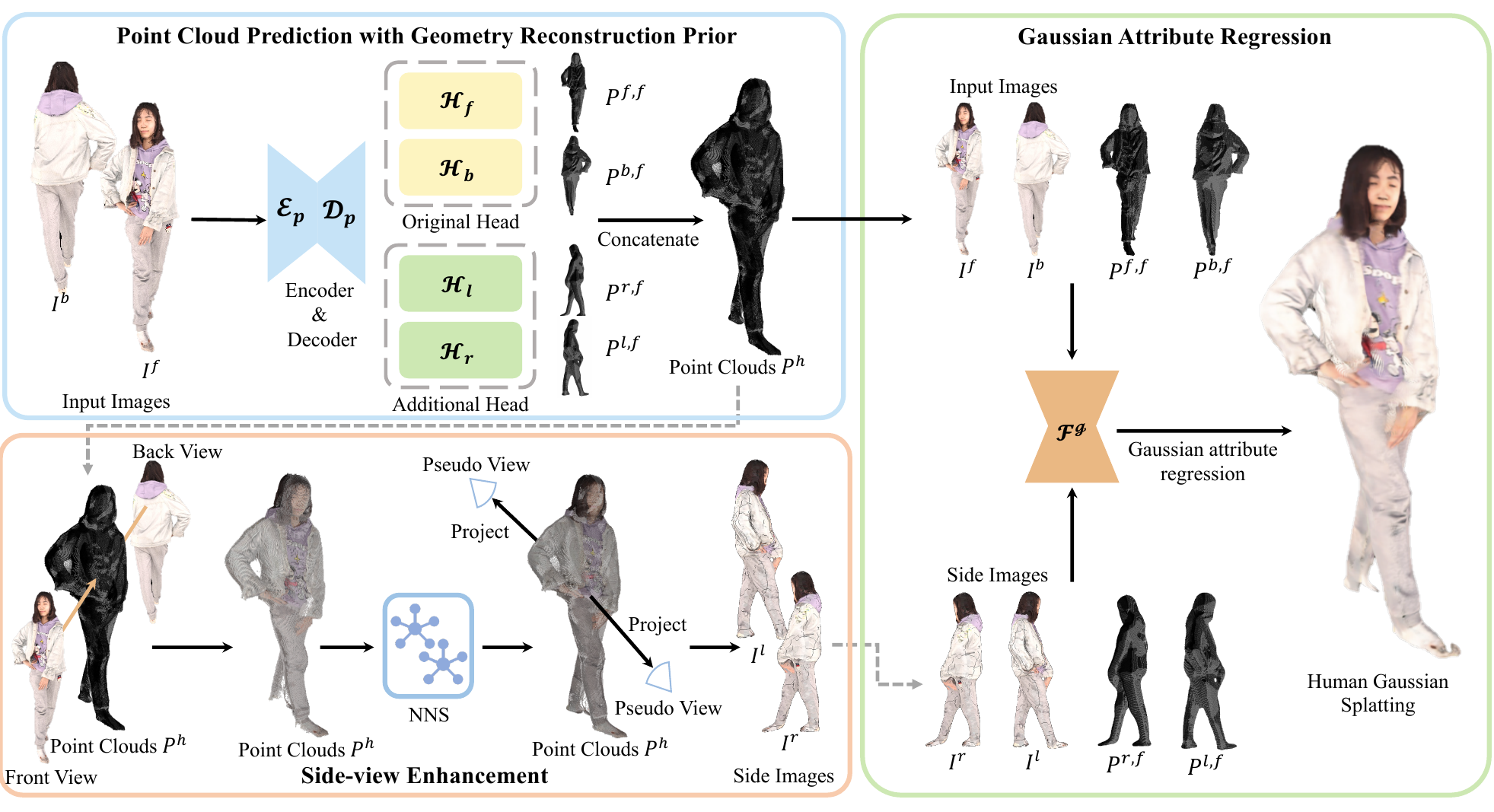}
    \caption{The framework of Snap-Snap. With the input front and back view images $I^{f}$ and $I^{b}$, the point cloud prediction model $\mathcal{R}_p$ generate the human point clouds from the front $P^{f,f}$, back $P^{b,f}$, left $P^{l,f}$, and right $P^{r,f}$ views. 
    Side-view color information is supplied by the side-view enhancement module. With the enhanced images $I^{l}, I^{r}$ and the input images $I^{f}, I^{b}$, we obtain fianl human Gaussians through Gaussian attribute regression $\mathcal{F}^{g}$.}
    \label{fig: pipeline}
\end{figure*}

\subsection{Preliminary}
\label{sec:preliminary}

\paragraph{3D Gaussian Splatting}
3D Gaussians~\cite{kerbl20233d} exhibit high quality during rendering while enjoying the speed of real-time rendering. 3D Gaussians represent space as ellipses of different sizes and orientations. Specifically, the 3D Gaussian is defined with its center position $\mu \in \mathbb{R}^3$, color $c \in \mathbb{R}^3$, opacity $o \in \mathbb{R}^1$, and covariance $\Sigma$. For optimization purposes, the covariance is decomposed into the scale $s \in \mathbb{R}^3$ and a quaternion $q \in \mathbb{R}^4$ representing rotation. Therefore, 3D Gaussians can be represented as:
\begin{equation}
    \label{eq: def3dgs}
    \theta=(\mu, c, o, \Sigma=(s, q))
\end{equation}
During the projection of 3D Gaussians onto the 2D pixel plane, 3D Gaussians accumulate along the corresponding rays. The entire rendering process is differentiable, which lays the foundation for optimization.

\subsection{Overview}

Given only the front and back RGB images of human body, we aim to obtain high-quality human Gaussian in a feed-forward manner without camera parameters.
Our generalizable human reconstruction algorithm can be divided into three stages as follows, as shown in Fig.~\ref{fig: pipeline}:

    \textbf{Point Cloud Prediction.}
    We redesign a human point cloud prediction model $\mathcal{R}_p$ to reconstruct complete human point clouds from four viewpoints (front, back, left and right viewpoints). We additionally introduce two heads to predict point clouds from side views under the condition of missing side geometry information, while maintaining alignment with the real-world coordinate system. By training $\mathcal{R}_p$ on human datasets, we adapt the geometric reconstruction prior to better fit the human domain. Finally, we concatenate the point clouds from all four views to form the complete human point clouds.
    
    \textbf{Side-view Enhancement.}
    Since the point cloud prediction model only predicts geometry for side views without color information, we construct an enhancement module to improve the left and right side views ${I^l, I^r}$ with the the front and back images ${I^f, I^b}$ in absence of camera parameters, thereby enhancing the completeness of the final representation.
    
    \textbf{Gaussian Attribute Regression.}
    Similar to point clouds prediction, we regress Gaussian attributes from four viewpoints directly. We input the point clouds of the four perspectives, the front and back images of the input, and the enhanced pseudo-color information into the Gaussian Regression network to obtain the final complete human Gaussian.

\subsection{Point Cloud Prediction}
\label{Point Cloud Prediction}

\begin{figure}[thbp]
    \centering
    \includegraphics[width=\columnwidth]{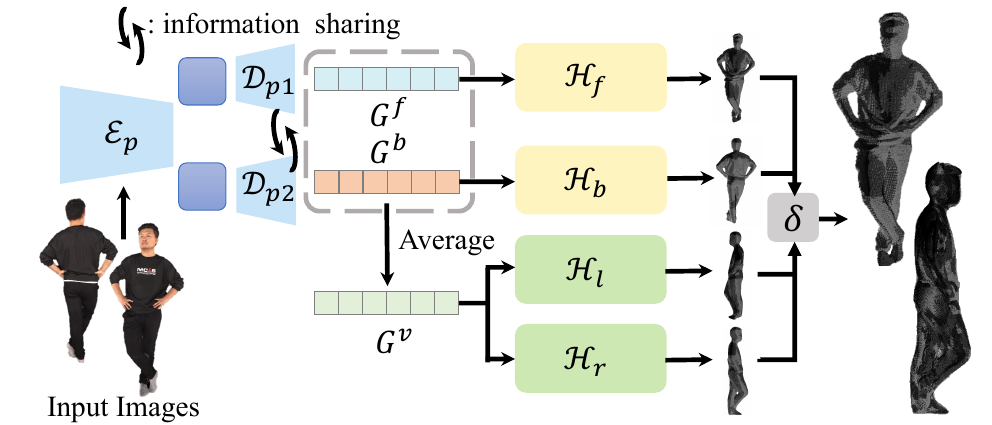}
    \caption{The framework of point cloud prediction network.}
    \label{fig: dust3r}
\end{figure}

We design the geometric reconstruction model $\mathcal{R}_p$ to perform point cloud prediction from four viewpoints: front, back, and two sides, ensuring the completeness of the predicted human point clouds. Specifically, we take the front and back images $I^f, I^b \in {\mathbb{R}}^{W \times W \times 3}$ as input, and process them through an encoder $\mathcal{E}_p$ and a decoder $\mathcal{D}_p$ with $B$ blocks to obtain intermediate image representations $G^f$ and $G^b$.

To fully leverage the priors from the foundation geometric reconstruction model, we adopt a similar architecture to DUSt3R~\cite{dust3r_cvpr24}. In order to generate point clouds from the front and back views, two prediction heads $\mathcal{H}_f, \mathcal{H}_b$ are used to process $G^f$ and $G^b$ respectively. Benefiting from the information exchange between front and back tokens within the decoder, we simply aggregate $G^f$ and $G^b$ to form the input tokens $G^v$ for the side-view heads $\mathcal{H}_l$ and $\mathcal{H}_r$.
By training the foundation geometric reconstruction model $\mathcal{R}_p$ on human data, the model learns to infer plausible geometry for two sides, even in the absence of explicit side-view observations.
\begin{equation}
\begin{split}
\{G^v\}_{i=1}^B = \left\{(G^f + G^b)/2 \right\}_{i=1}^B, \\
P^{l,f} = \mathcal{H}_l\left(\{G^v\}_{i=1}^B\right),
P^{r,f} = \mathcal{H}_r\left(\{G^v\}_{i=1}^B\right),
\end{split}
\label{eq:points2}
\end{equation}

We concatenate the point clouds of the front, back, left, and right in the perspective of the front viewpoint to obtain the final complete point clouds of the human body. Furthermore, to align the predicted point cloud with the human in the real world, we introduce a learnable parameter to estimate the actual human scale, thus obtaining a scaling factor $\delta$ and generate the final human point cloud with the correct proportions in the real-world coordinate.
\begin{equation}
    \begin{split}
    P^{h} = \delta * (P^{f,f} \oplus P^{b,f}\oplus P^{l,f}\oplus P^{r,f}),
    \end{split}
\end{equation}

Based on the priors of the foundation geometry reconstruction model and training on human datasets, we obtain the complete human point clouds from four views predicted in the perspective of the front viewpoint, even with almost no overlapping input. $\mathcal{R}_p$ implicitly learns the mapping relationships between viewpoints, directly learning the positional relationships of point clouds between different viewpoints, enabling the acquisition of the complete human point clouds even without camera parameters.

\subsection{Side-view Enhancement}
\label{Side-view Enhancement}

Since we only use the images from the front and back views, the predicted side point clouds $P^{l,f}, P^{r,f}$ do not contain any color information, leading to a severe lack of color information on the side views of the human body. To address this, we propose a simple nearest neighbor search (NNS) algorithm to warp the color information from the front and back views to the side views.

In Sec.~\ref{Point Cloud Prediction}, the human front and back point clouds $P^{f,f}, P^{b,f}$ along with side point clouds $P^{l,f}, P^{r,f}$ are obtained, but color information is missing. Since the point clouds from the geometric reconstruction model $\mathcal{R}_p$ is pixel-wise, we can easily obtain the color information $C^{f,f}, C^{b,f}$ of front and back point clouds $P^{f,f}, P^{b,f}$. Specifically, we project the front and back image pixels to their corresponding point clouds. 

Due to the missing color information on the sides, the final reconstruction results are adversely affected. The NNS algorithm is further used to transfer colors in front and back point clouds to side point clouds, which are then unprojected to side pseudo-views to obtain the side color information. We use NNS to find the nearest neighbors of the side point clouds $P^{l,f}, P^{r,f}$ in the known colored point clouds $P^{f,f}, P^{b,f}$, and then assign their color information to the side point clouds, thereby obtaining the color information $C^{l,f}$ and $C^{r,f}$ for the side point clouds. Taking the left view for example. Assume $P^{l,f} = \{p^{l,f}_1, p^{l,f}_2, \dots, p^{l,f}_n\}, C^{l,f} = \{c^{l,f}_1, c^{l,f}_2, \dots, c^{l,f}_n\}$, this process can be formulated as
\begin{equation} 
\begin{aligned} 
\label{eq: com points} 
\forall  i \in \{ 1,2,...n \}: \\
&j = F_\text{nns}(\{P^{f,f}, P^{b,f}\}; p^{l,f}_i),\\
& c^{l,f}_i \leftarrow \text{Index}( \{C^{f,f}, C^{b,f}\};j),
\end{aligned} 
\end{equation}
where $F_{nns}$ denotes the process of searching the nearest neighbor index $j$ in the point cloud set $\{P^{f,f}, P^{b,f}\}$ for each element $p^{l,f}_i \in P^{l,f}$. 
$Index$ denotes an indexing process which looks up the color set $\{C^{f,f}, C^{b,f}\}$ using index $j$ and then assigns the fetched value to $c^{l,f}_i$.
Similarly, we can obtain the color information for $P^{r,f}$.
Since $P^{l,f}$ and $P^{r,f}$ are also predicted in a pixel-wise manner, we can easily establish correspondence between the point cloud colors $C^{l,f}, C^{r,f}$ and the corresponding side pseudo-view pixels $I^l, I^r$ without known camera parameters.

\begin{table*}[t]
\centering
\small
\setlength{\tabcolsep}{2pt}
\resizebox{\textwidth}{!}{
\begin{tabular}{lcccccccccccc}
 \toprule
  & & & &\multicolumn{3}{c}{\textbf{Thuman2.0~\cite{yu2021function4d}}} & \multicolumn{3}{c}{\textbf{2K2K~\cite{han20232k2k}}} & \multicolumn{3}{c}{\textbf{4D-Dress~\cite{4d_dress}}} \\
 \cmidrule(lr){5-7} \cmidrule(lr){8-10} \cmidrule(lr){11-13}
 Method & Views & SMPL-X~\cite{smplx} & Infer. (ms) &\multicolumn{1}{c}{PSNR$\uparrow$} & \multicolumn{1}{c}{SSIM $\uparrow$} & \multicolumn{1}{c}{LPIPS $\downarrow$} & \multicolumn{1}{c}{PSNR$\uparrow$} & \multicolumn{1}{c}{SSIM$\uparrow$} & \multicolumn{1}{c}{LPIPS $\downarrow$} & \multicolumn{1}{c}{PSNR$\uparrow$} & \multicolumn{1}{c}{SSIM$\uparrow$} & \multicolumn{1}{c}{LPIPS $\downarrow$} \\
 \toprule
 GPS-Gaussian~\cite{zheng2024gps} & 5 & N/A & 144 & 20.70 & 88.11 & 19.15 & 21.68 & \textbf{88.45} & 17.17 & 20.47 & 84.78 & 22.43 \\
 GHG~\cite{ghg_2024} & 3 & GT & 2858 & 21.99 & 88.45 & 13.79 & 21.73 & 87.50 & 13.88 & 21.20 & \textbf{86.03} & 17.24 \\
 GHG & 2 & GT & 2853 & 21.79 & 87.66 & 17.54 & 21.26 & 86.38 & 17.54 & 20.69 & 85.13 & 20.58 \\
 \toprule
 GHG & 2 & Estimated & 10696 & 16.71 & 81.92 & 26.22 & 16.72 & 80.48 & 26.53 & 18.34 & 80.99 & 26.55 \\
\textbf{Snap-Snap (Ours)} & 2 & N/A & 190 & \textbf{22.44
} & \textbf{88.78} & \textbf{13.24}  & \textbf{22.38} & 88.01 & \textbf{13.08} & \textbf{21.62} & 85.55 & \textbf{17.03}  \\
\bottomrule
\end{tabular}
}
\caption{ We present the comparison results with the GPS-Gaussian~\cite{zheng2024gps} and GHG~\cite{ghg_2024}. We find that GHG uses the ground-truth SMPL-X~\cite{smplx} parameters during inference. For a fair comparison, we estimate the SMPL-X parameters only using two viewpoints though EasyMocap~\cite{easymocap}. Infer. denotes the total inference time from receiving images and masks to the final Gaussians. The detailed time-consuming analysis is presented in the supplementary materials.}
\label{tab: com}
\end{table*}

\subsection{Gaussian Attribute Regression}
In a feed-forward manner, we predict 3D Gaussians $\theta=(\mu, c, o, \Sigma=(s, q))$ based on the obtained human point clouds. 
The point cloud prior predicted by Sec.~\ref{Point Cloud Prediction} represents points upon the human body surface. Considering the differences between point clouds and 3D Gaussians, the prediction of absolute 3D coordinates $\mu$ is reformulated as predicting the offset $\Delta \mu$ relative to the point cloud prior. In addition, we need to predict the scale, opacity, color, and rotation of the Gaussians.
To reconstruct the entire human body, we regress the Gaussian attributes from the input front and back viewpoints and the left and right pseudo-viewpoints. With the point clouds predicted from the four viewpoints $P^{f,f}, P^{b,f}, P^{l,f}, P^{r,f} \in {\mathbb{R}}^{W \times W \times 3}$ and the color maps $I^f, I^b, I^{l}, I^{r} \in {\mathbb{R}}^{W \times W \times 3}$, we use a UNet-like~\cite{ronneberger2015u} network $\mathcal{F}^{g}$ to predict the Gaussian attributes $\theta=\{\Delta \mu, c, o, s, q\}$:

\begin{equation}
\begin{aligned}
    \theta_f, \theta_b  &= \mathcal{F}^{g} ( \{P^{f,f}, I^f\}, \{P^{b,f}, I^b\}), \\
    \theta_l,  \theta_r &= \mathcal{F}^{g} (\{P^{l,f}, I^l\}, \{P^{r,f}, I^r\} ), \\
    \theta &= \theta_f \oplus \theta_b \oplus \theta_l \oplus \theta_r ,
    \label{eq: gs_re}
\end{aligned}
\end{equation}
where $\theta_f, \theta_b, \theta_l, \theta_r$ denote the Gaussians predicted from front, back, left, and right viewpoints, respectively. By concatenating them together, we obtain the final Gaussian representation $\theta$ for the entire human body.

\subsection{Training and Inference}

The entire framework's learnable components can be divided into two modules: (1) The point cloud prediction network, and (2) The Gaussian regression network. To ensure the performance of the algorithm, we train the two modules separately.

\paragraph{Training Stage 1:} For the point cloud prediction network, we use 3D point clouds $P^{gt}$ and 2D image mask $M^{gt}$ as supervision. The L2 loss and cross-entropy loss are denoted as the regression loss $\mathcal{L}_{reg}$ and the confidence loss $\mathcal{L}_{conf}$.
\begin{equation}
    \mathcal{L}_{stage1} = {L}_{reg}\left(P^{h}, P^{gt}\right) + {L}_{conf}(M^{conf},M^{gt}).
\end{equation}

\paragraph{Training Stage 2:} For the Gaussian regression network, we render image $I^{render}$ from novel views with differentiable splatting, and use the ground-truth image $I^{gt}$ as training supervision. L1 loss and SSIM loss, denoted as $\mathcal{L}_{rgb}$ and $\mathcal{L}_{ssim}$, are employed to train the Gaussian regression.
\begin{equation}
\label{eq: stage 2 loss} 
    \resizebox{.9\hsize}{!}{$
    \mathcal{L}_{stage2} = \beta {L}_{rgb}\left(I^{render}, I^{gt}\right) + (1-\beta) \mathcal{L}_{ssim}\left(I^{render}, I^{gt}\right).
    $}
\end{equation}
\paragraph{Inference:} During inference, we first use the input front and back view images $I^{f}, I^{b}$ to obtain the point clouds of the human body $P^{f,f}, P^{b,f}, P^{l,f}, P^{t,f}$ through the point cloud prediction model $\mathcal{R}_p$. Then, we obtain the enhanced images $I^{l}, I^{r}$ through the side-view enhancement algorithm. Based on the enhanced images $I^{l}, I^{r}$ and the input images $I^{f}, I^{b}$ and the point clouds of the human body $P^{f,f}, P^{b,f}, P^{l,f}, P^{r,f}$, the human Gaussian are obtained through Gaussian attribute regression $\mathcal{F}^{g}$.

\begin{figure*}[t]
    \centering
    \includegraphics[width=1\linewidth]{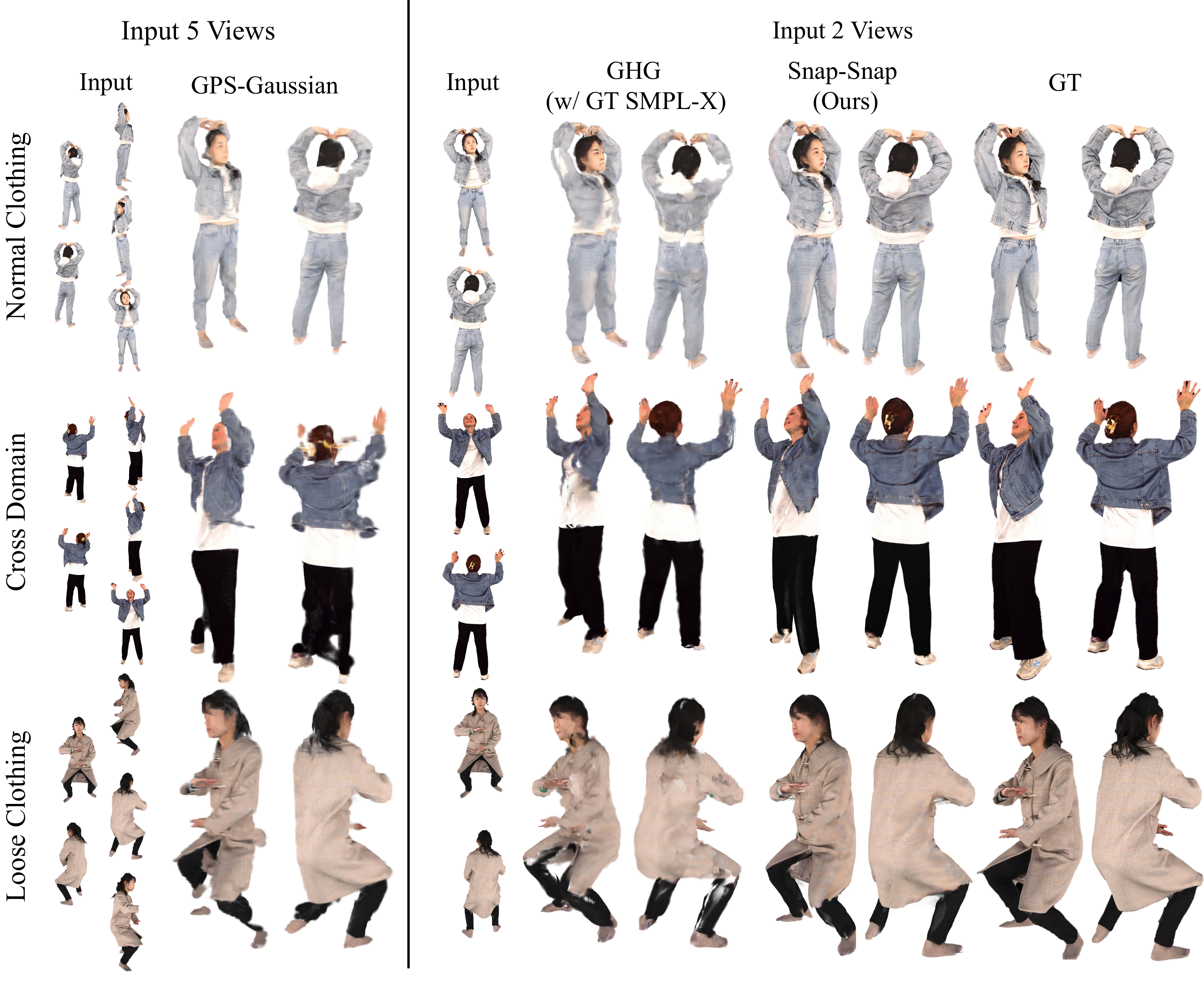}
    \vspace{-20pt}
    \caption{Visual comparisons with GPS-Gaussian~\cite{zheng2024gps} and GHG~\cite{ghg_2024}.}
    \label{fig: vis}
\end{figure*}

\section{Experiments}

\subsection{Implementation Details}

We train our model on a single RTX 4090 with 24G memory and set the batch size as 1. For the training of each module, we apply the AdamW ~\cite{loshchilov2017decoupled} optimizer with a learning rate of $1\times 10^{-4}$, and a weight decay of $5\times 10^{-2}$. The learning rate is cosine annealed to $1\times 10^{-7}$ during the training. The training iterations for stage 1 are $100k$, while the training iterations for stages 2 are only $50k$. The training times for the two stages are approximately 13 and 6 hours respectively. Besides, the $\beta$ in Eq.~\ref{eq: stage 2 loss} for the stage 2 is 0.8.

\paragraph{Datasets.}

We train and evaluate on the Thuman2.0 dataset~\cite{yu2021function4d} while evaluating on the 2K2K~\cite{han20232k2k} and 4D-Dress~\cite{4d_dress} for cross-domain evaluation. The THuman2.0 dataset consists of 526 high-quality human assets. Following \cite{ghg_2024}, we select the same 100 subjects to evaluate the algorithm. Due to the requirement of ground-truth SMPL-X parameters by GHG~\cite{ghg_2024}, we select the first 100 subjects from the 2K2K training set for evaluation.
Furthermore, to evaluate the reconstruction quality of loose clothing, we select 50 loose clothing examples from Thuman2.1~\cite{yu2021function4d} (exclude human scans of Thuman2.0) by calculating the chamfer distance between the SMPL-X~\cite{smplx} mesh and the ground-truth human meshes. We sort the Chamfer Distance from largest to smallest and select the top human bodies as Loose Clothes val set.

\subsection{Comparisons}

In Tab.~\ref{tab: com}, we present the comparison results with GPS-Gaussian~\cite{zheng2024gps} and GHG~\cite{ghg_2024} with three metrics: Peak Signal-to-Noise Ratio (PSNR), Structure Similarity Index Measure (SSIM)~\cite{wang2004image}, and Learned Perceptual Image Patch Similarity (LPIPS)~\cite{zhang2018unreasonable} to assess the quality of the rendered images. LPIPS is calculated with AlexNet~\cite{krizhevsky2012imagenet} model. The resolution of the rendered images is 1024 $\times$ 1024, and we only use the human regions given by the bounding box of humans when calculating the metrics following the GHG. For better comparison, we provide visual results in Fig.~\ref{fig: vis}, where the resolution of the rendered images for all methods is 2024 $\times$ 2048.
GHG under 2 views is trained by the publicly released code, while we use the released inpainting model weight. GPS-Gaussian under 5 views is completely trained following the instructions.

\begin{table}[t]
\centering
\small
\setlength{\tabcolsep}{2pt}
\begin{tabular}{lccccc}
 \toprule
 Method & Views & SMPL-X~\cite{smplx} & PSNR$\uparrow$ & SSIM $\uparrow$ & LPIPS $\downarrow$ \\
 \toprule
 GPS-Gaussian~\cite{zheng2024gps} & 5 & N/A & 19.49 & \textbf{84.45} & 22.02 \\
 GHG~\cite{ghg_2024} & 3 & GT & 20.02 & 84.36 & 17.67 \\
 GHG & 2 & GT & 19.74 & 83.44 & 21.60 \\
 \toprule
 GHG & 2 & Estimated & 15.80 & 78.77 & 29.08 \\
\textbf{Snap-Snap (Ours)} & 2 & N/A & \textbf{20.98} & 84.42 & \textbf{17.08} \\
\bottomrule
\end{tabular}
\caption{The comparison on loose clothes val set.}
\label{tab: loose}
\end{table}

\begin{table}[t]
\centering
\small
\setlength{\tabcolsep}{2pt}
\begin{tabular}{lccccc}
 \toprule
 Method & Views & SMPL-X & PSNR$\uparrow$ & SSIM$\uparrow$ & LPIPS $\downarrow$ \\
 \toprule
 SiTH~\cite{ho2024sith} & 2 & GT & 15.10 & 76.74 & 31.68 \\
\textbf{Snap-Snap (Ours)} & 2 & N/A & \textbf{22.44} & \textbf{88.78} & \textbf{13.24} \\
\bottomrule
\end{tabular}
\caption{The comparison between Snap-Snap and mesh-based human reconstruction SiTH~\cite{ho2024sith}.}
\label{tab: with_sith}
\end{table}

The reconstruction quality of GPS-Gaussian is lower than ours even with five viewpoints. In Fig.~\ref{fig: vis} we can see that GPS-Gaussian's reconstruction shows missing body parts due to the limitations of the depth estimation module in GPS-Gaussian, which cannot provide reasonable results due to the sparsity of the viewpoints. Unlike GPS-Gaussian, our method completes the side views both in terms of point clouds and colors, which can directly infer the complete human point clouds from the front and back views. 

GHG completes human body from sparse viewpoints based on the human prior SMPL-X. We find that GHG uses the ground-truth SMPL-X parameters during the inference, which are calculated from far more than two viewpoints. To fairly evaluate GHG's reconstruction quality, we use EasyMocap~\cite{easymocap} to estimate the SMPL-X parameters from two viewpoints.
Moreover, our method enables human reconstruction at the millisecond level shown in Tab.~\ref{tab: com}, in contrast to GHG which requires time computing SMPL-X parameters and preprocessing data.

In Tab.~\ref{tab: loose}, we evaluate our method on human bodies with loose clothing, demonstrating the robustness of our method over different human bodies. In the visual results, we can see that due to GHG being based on SMPL-X, it cannot reconstruct loose clothing well. Our method directly infers complete geometric information through the point cloud prediction model, achieving good modeling even for loose clothing.

\begin{figure}[t]
    \centering
    \includegraphics[width=1\linewidth]{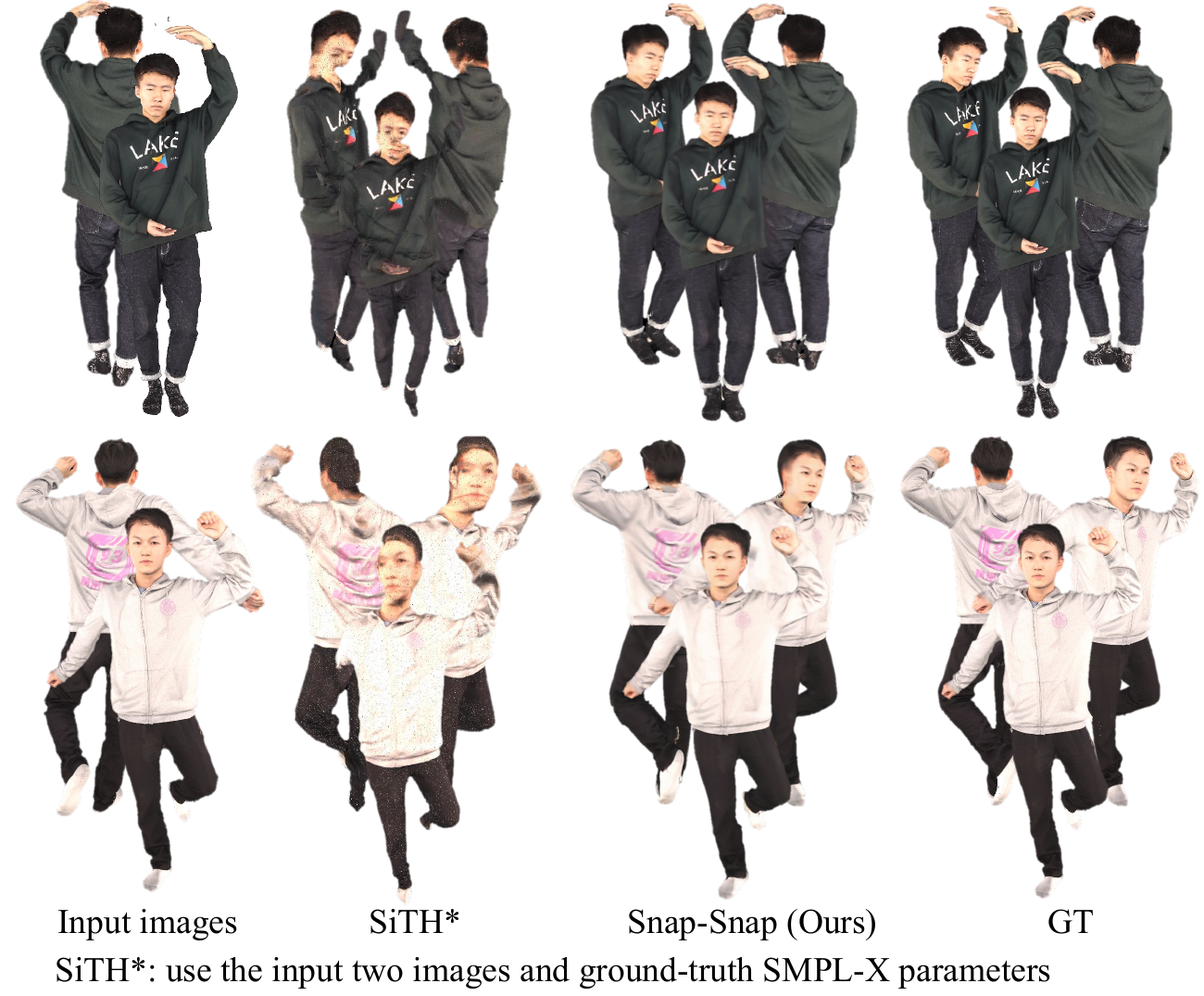}
    \caption{Qualitative comparisons with SiTH~\cite{ho2024sith}.}
    \label{fig: with_sith}
\end{figure}

\begin{figure}[t]
    \centering
    \includegraphics[width=\linewidth]{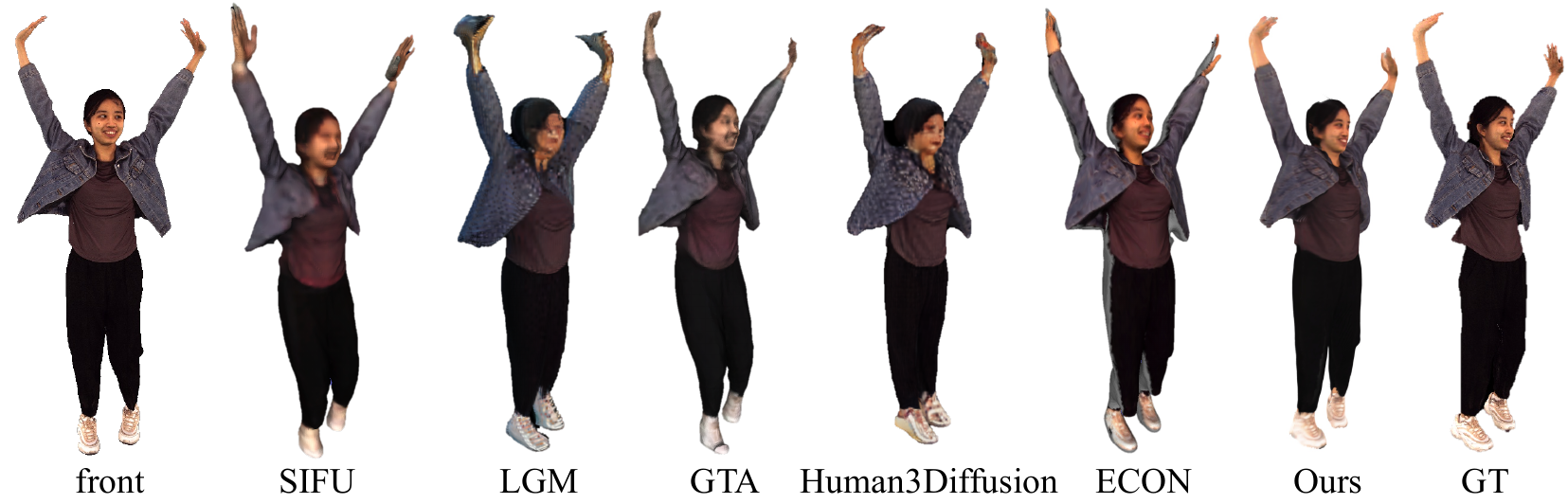}
    \caption{Visual comparisons of Snap-Snap with single-view reconstruction methods.}
    \label{fig: singe_view comparsion}
\end{figure}

\begin{figure}[t]
    \centering
    \includegraphics[width=1\linewidth]{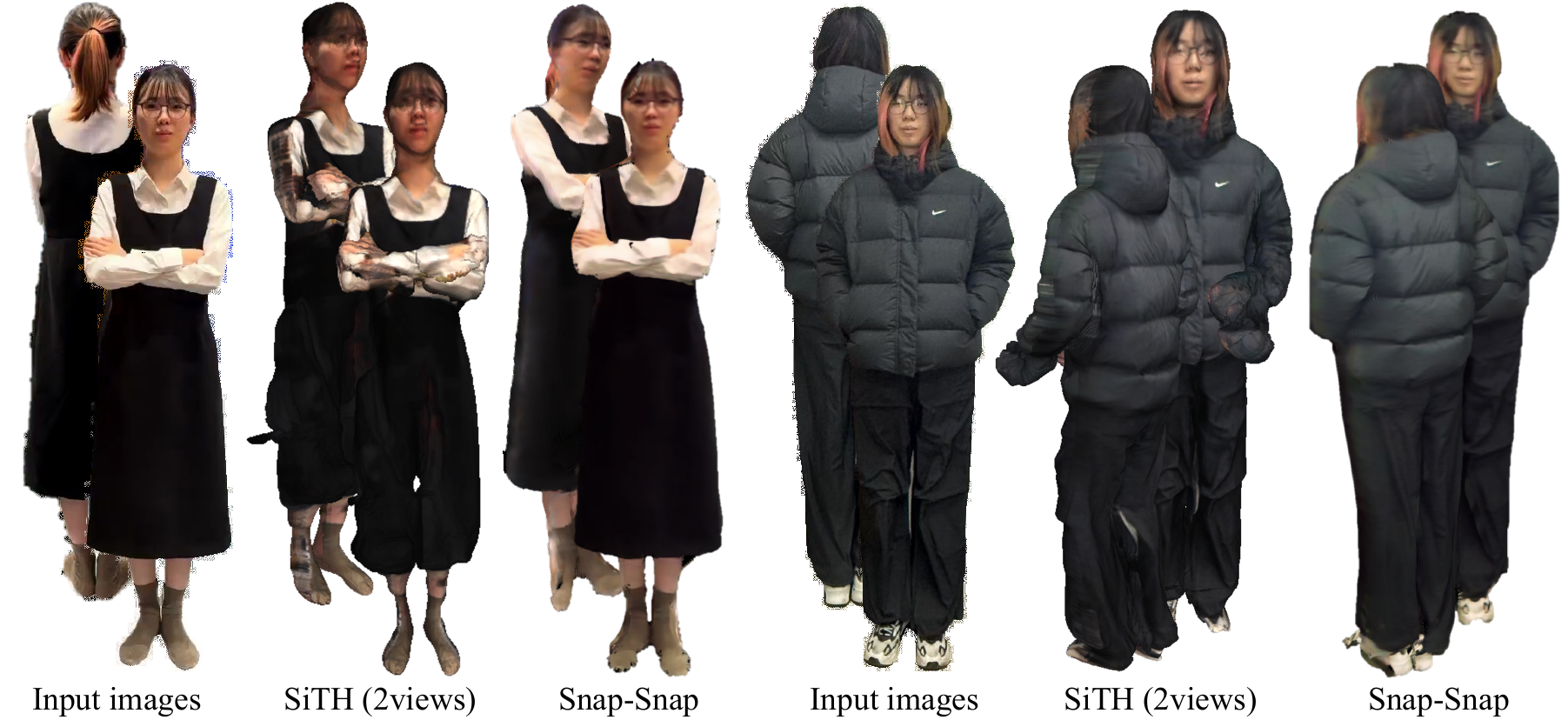}
    \caption{Reconstruction results from in-the-wild data.}
    \label{fig: wild}
\end{figure}

\begin{figure}[t]
    \centering
    \includegraphics[width=1\linewidth]{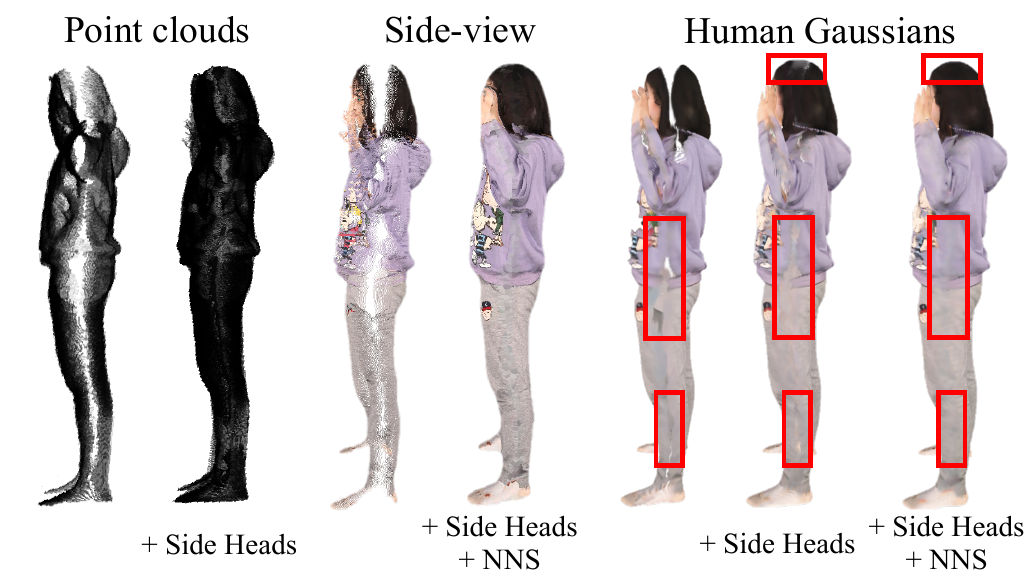}
    \caption{We present the visual results of ablation studies on the additional side-view head in the point cloud prediction network and the side-view enhancement module.}
    \label{fig: abla}
\end{figure}

\begin{table}[t]
\centering
\small
\begin{tabular}{ccccc}
\toprule
Side Head & NNS & PSNR $\uparrow$ & SSIM $\uparrow$ &  \begin{tabular}{c}LPIPS$\downarrow$ \\\end{tabular} \\
\midrule
 $\times$ & $\times$ & 22.15 & 88.20 & 14.11 \\
 $\checkmark$ & $\times$ & 22.34 & 88.39 & 13.75 \\
 $\checkmark$ & $\checkmark$ & \textbf{22.44} & \textbf{88.78} & \textbf{12.47} \\
\bottomrule
\end{tabular}
\caption{We perform ablation studies on the additional side-view heads in the point cloud prediction network and the side-view enhancement module.}
\label{tab: head and inpainting}
\end{table}

\subsection{Comparison with Single-view Reconstruction}
\label{misalignment}
To further evaluate the effectiveness of our method, we compare it with the mesh-based single-view reconstruction approaches. To minimize the influence of generative components~\cite{ho2024sith} on the human reconstruction, we select SiTH~\cite{ho2024sith} as the method for quantitative comparison. SiTH allows replacing the pseudo back-view with the ground-truth back-view image, and we use the ground-truth SMPL-X parameters as the input human prior.

Since mesh-based human reconstruction methods are often not aligned with the world coordinate system, we apply ICP registration ~\cite{rusinkiewicz2001icp} using the ground-truth mesh and render the registered meshes for evaluation. We compute PSNR on the rendered images, as shown in Tab.~\ref{tab: with_sith}, with more qualitative results provided in Fig.~\ref{fig: with_sith}. The results demonstrate that our method achieves better reconstruction quality and aligns more accurately with the human pose in the input images. As shown in Fig.~\ref{fig: with_sith}, despite using accurate human prior (SMPL-X parameters), SiTH still produces distorted poses. In contrast, our method achieves high consistency with the target human in both pose and texture.
Furthermore, we observe that due to the misalignment between mesh-based reconstruction results and the real-world coordinate system, evaluating such methods using PSNR can be unreliable.

Considering the alignment issue mentioned, we further provide a qualitative comparison with existing single-view reconstruction methods~\cite{xue2024human3diffusion,xiu2023econ,zhang2023gta,zhang2024sifu,tang2025lgm} in Fig.~\ref{fig: singe_view comparsion}. These single-image methods exhibit significantly worse reconstruction consistency than ours, even when manually aligned. Moreover, the results of single-view reconstruction are often difficult to align with the human body scale in the world coordinate system, whereas our reconstructions maintain a reasonable scale due to the point clouds prediction network.

\subsection{Reconstruction from In-the-Wild Data}
To validate our method on capture from low-cost mobile phone, we use two mobile phones to build a capture setup and reconstruct human bodies from the collected data. The reconstruction results are shown in Fig.~\ref{fig: wild}. Details about the low-cost data collection device are provided in the supplementary material.
In the absence of camera parameters under the simple capture setup, GPS-Gaussian and GHG are unable to successfully reconstruct human. Consequently, we provide a qualitative comparison with the two-view SiTH. Our method achieves superior human reconstruction in a more robust and end-to-end manner.

\subsection{Ablation Studies}

We conduct ablation experiments on the heads of the point cloud prediction model $\mathcal{R}_p$ and the side-view enhancement algorithm to verify the impact of these two modules on the human reconstruction, as shown in Tab.~\ref{tab: head and inpainting}. More ablation studies are given in the supplementary material.

\paragraph{Heads of Point Cloud Prediction Model.}
We perform ablation experiments on the side heads in the point cloud prediction model $\mathcal{R}_p$, as shown in the  Tab.~\ref{tab: head and inpainting}. We find that the point clouds of the human exhibit obvious gaps without the side heads, adversely affecting the final human reconstruction performance. To compensate for these gaps, we use two heads to predict the side point clouds of the human body, which greatly improves the completion of the point clouds and the final reconstruction quality shown in Fig.~\ref{fig: abla}.

\paragraph{Side-view Enhancement.}
We conduct ablation experiments on the side-view enhancement algorithm shown in Tab.~\ref{tab: head and inpainting}. Instead of using NNS to obtain side-view images, we directly feed the front and back images into the Gaussian attribute regression under the condition that camera parameters are not available. By leveraging the color warping of NNS which based on the spatial relationships within the point clouds, our side-enhancement algorithm achieves superior reconstruction results. Fig.~\ref{fig: abla} illustrates the human point clouds with colors after wrapping. It can be observed that the side-view enhancement leads to more consistent textures, particularly on the side views.

\subsection{Scalability}

\begin{table}[t]
\centering
\begin{tabular}{c|ccc}
\toprule
Training Scans & PSNR $\uparrow$ & SSIM $\uparrow$ & LPIPS $\downarrow$\\
\midrule
 426 & 22.44 & 88.78 & \textbf{13.47} \\
 2992 & \textbf{22.77} & \textbf{88.98} & 13.56 \\
\bottomrule
\end{tabular}
\caption{Evaluation results with more datasets.}
\label{tab: Scale Up}
\end{table}

To further explore the relationship between the current method and the amount of training data, we expand the dataset by adding Thuman2.1~\cite{yu2021function4d} (1919 scans) and the CustomHuman dataset~\cite{ho2023custom} (647 scans) to Thuman2.0~\cite{yu2021function4d} (426 scans), resulting in a total of 2992 human scans for training. Experimental results in Tab.~\ref{tab: Scale Up} show that as the training data increases, the reconstruction performance of the current method further improves, demonstrating its strong scalability.

\section{Conclusion}
In this paper, we propose a feed-forward framework capable of directly predicting 3D human Gaussians from only two images in 190 ms on one GPU.
We redesign a geometric reconstruction model for human point clouds prediction from four viewpoints, adapting the robust geometric prior from recent foundational reconstruction models to the specific human domain via training on human data. To complete the omitted information from the input two images, we propose a simple nearest neighbor search algorithm. 3D human Gaussians can be directly obtained from the completed point clouds. We expect our method could widen the applications of human body reconstruction.


{
    \small
    \bibliographystyle{ieeenat_fullname}
    \bibliography{main}
}

\appendix

\section{Appendix}

\begin{figure}[t]
    \centering
    \includegraphics[width=\linewidth]{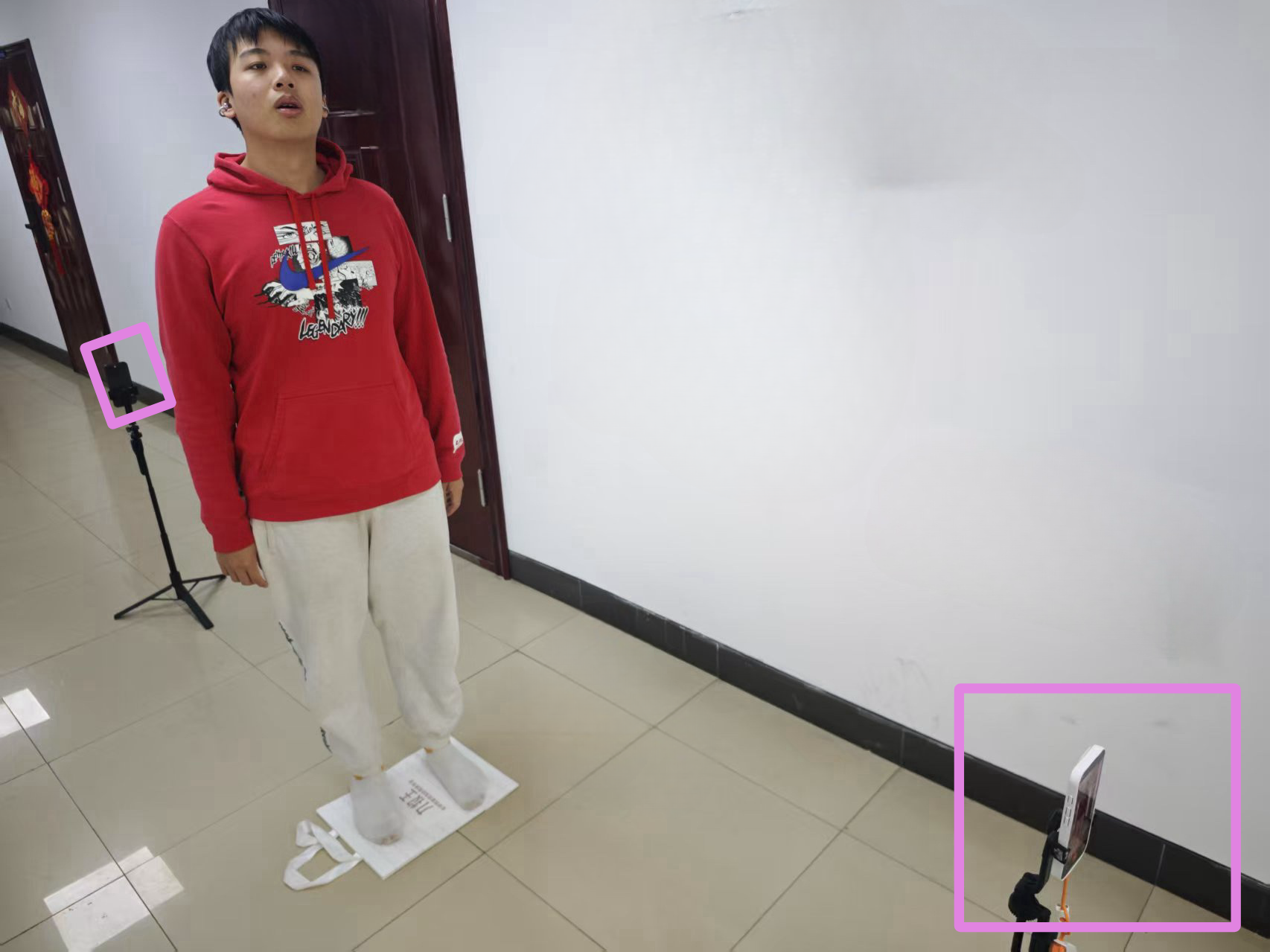}
    \caption{Data collection setup visualization.}
    \label{fig: Collection}
\end{figure}

\subsection{Data Collection Setup Visualization}
We present the low-cost data collection setup in Fig.~\ref{fig: Collection}. We place the two phones on two stands and synchronize the images captured by the phones with a computer. Directly reconstructing a human body from images taken in random poses is quite challenging, so to make the camera poses of the two mobile phones as similar as possible to those in the training set, we manually adjust the poses of the two phones to reduce the reconstruction difficulty.

\subsection{Visual Comparison with Two-view 3D Generation}

\begin{figure}[t]
    \centering
    \includegraphics[width=\linewidth]{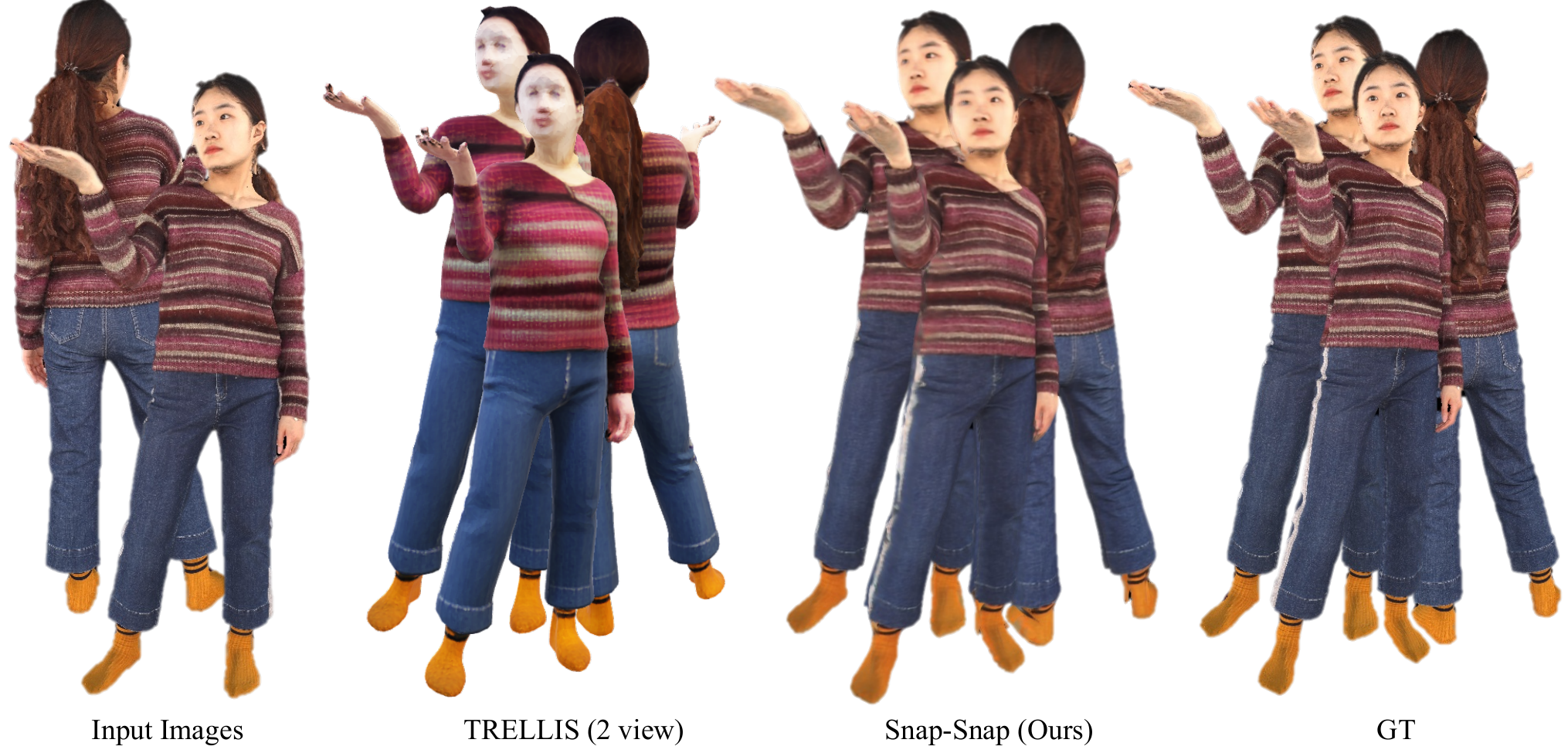}
    \caption{Visual comparisons of Snap-Snap with TRELLIS~\cite{xiang2025trellis}.}
    \label{fig: with_trellis}
\end{figure}

We provide a visual comparison between our method and TRELLIS~\cite{xiang2025trellis}, which can reconstruct 3D assets based on two images (\emph{e.g.}, human bodies). As shown in Fig.~\ref{fig: with_trellis}, our method demonstrates superior clarity and texture consistency in a faster inference speed, especially in terms of skin tone.

\begin{figure}[t]
    \centering
    \includegraphics[width=\linewidth]{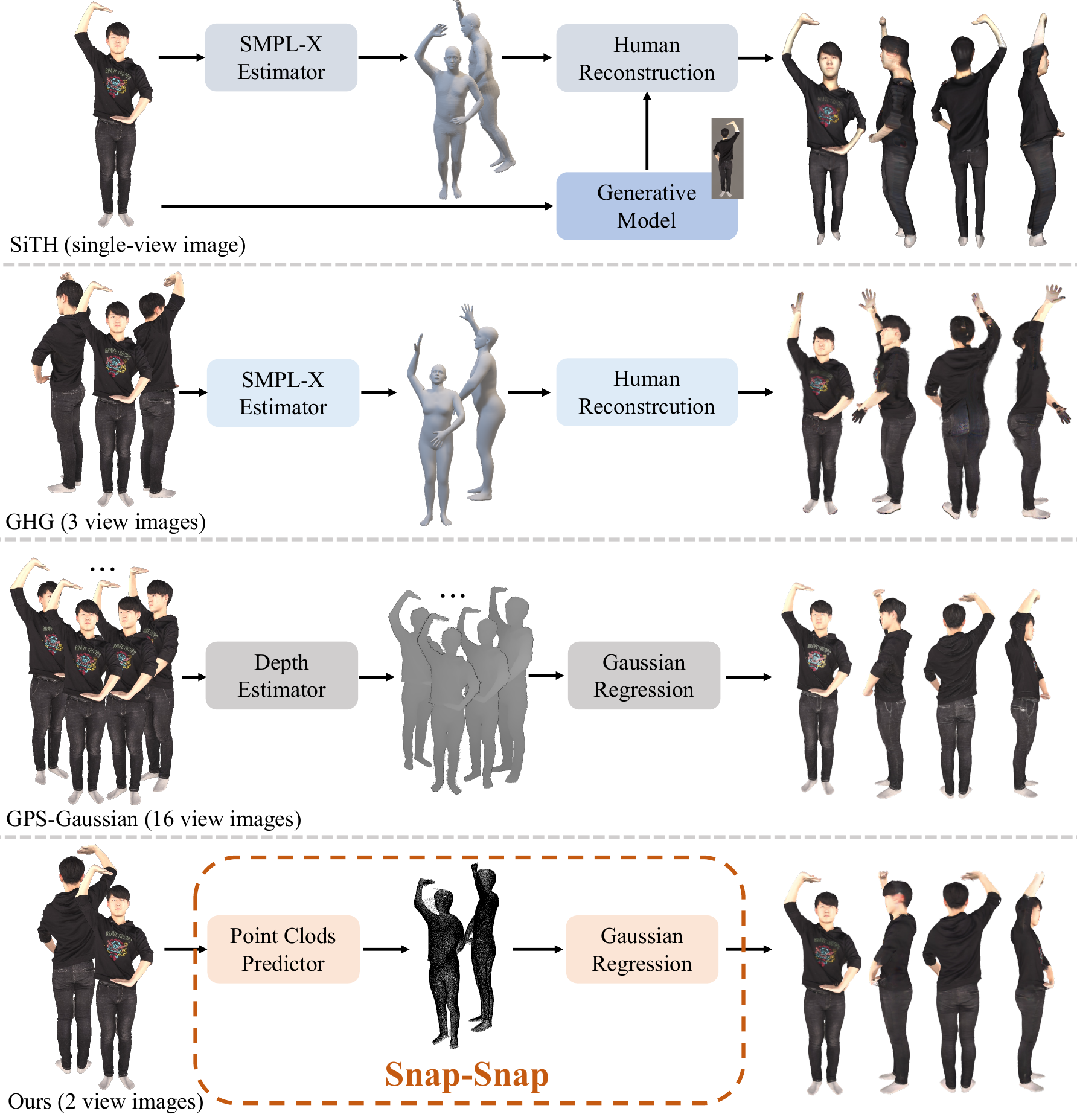}
    \caption{Comparison of pipelines with different methods.}
    \label{fig: singe_view comparsion}
\end{figure}

\subsection{Comparison with Different Methods}

To further highlight the advantages of our approach, we provide a comparative illustration of different representative frameworks: single-view reconstruction methods (\emph{e.g.}, SiTH~\cite{ho2024sith}), sparse-view methods based on SMPL-X estimations (\emph{e.g.}, GHG~\cite{ghg_2024}), and dense-view depth-based methods (\emph{e.g.}, GPS-Gaussian~\cite{zheng2024gps}). Unlike existing approaches, our method achieves high-quality human reconstruction in milliseconds, using only a minimal number of input views, without relying on generative models~\cite{rombach2022sith_diffusion} which often lead to uncontrollable textures and human prior estimations which often exist misalignment due to the lacking of views.

\begin{table}[H]
\centering
\begin{tabular}{c|ccc}
\toprule
Regression Networks & PSNR $\uparrow$ & SSIM $\uparrow$ & LPIPS $\downarrow$\\
\midrule
 2 & 22.41 & 88.74 & \textbf{13.36} \\
 1 (Ours) & \textbf{22.44} & \textbf{88.78} & 13.47\\
\bottomrule
\end{tabular}
\caption{We perform ablation studies on the number of Gaussian regressions networks.}
\label{tab: regression_head}
\end{table}

\begin{table}[H]
\centering
\begin{tabular}{c|cccc}
\toprule
Pretraining & PSNR $\uparrow$ & SSIM $\uparrow$ & LPIPS $\downarrow$ \\
\midrule
 $\times$ & 21.11 & 87.49 & 16.22 \\
 $\checkmark$ & \textbf{22.44} & \textbf{88.78} & \textbf{13.47}\\
\bottomrule
\end{tabular}
\caption{We perform ablation studies on the impact of DUSt3R~\cite{dust3r_cvpr24} pretraining weight.}
\label{tab: pretraining}
\end{table}

\subsection{Uncontrollability of Generative Models}

\begin{figure}[t]
    \centering
    \includegraphics[width=\linewidth]{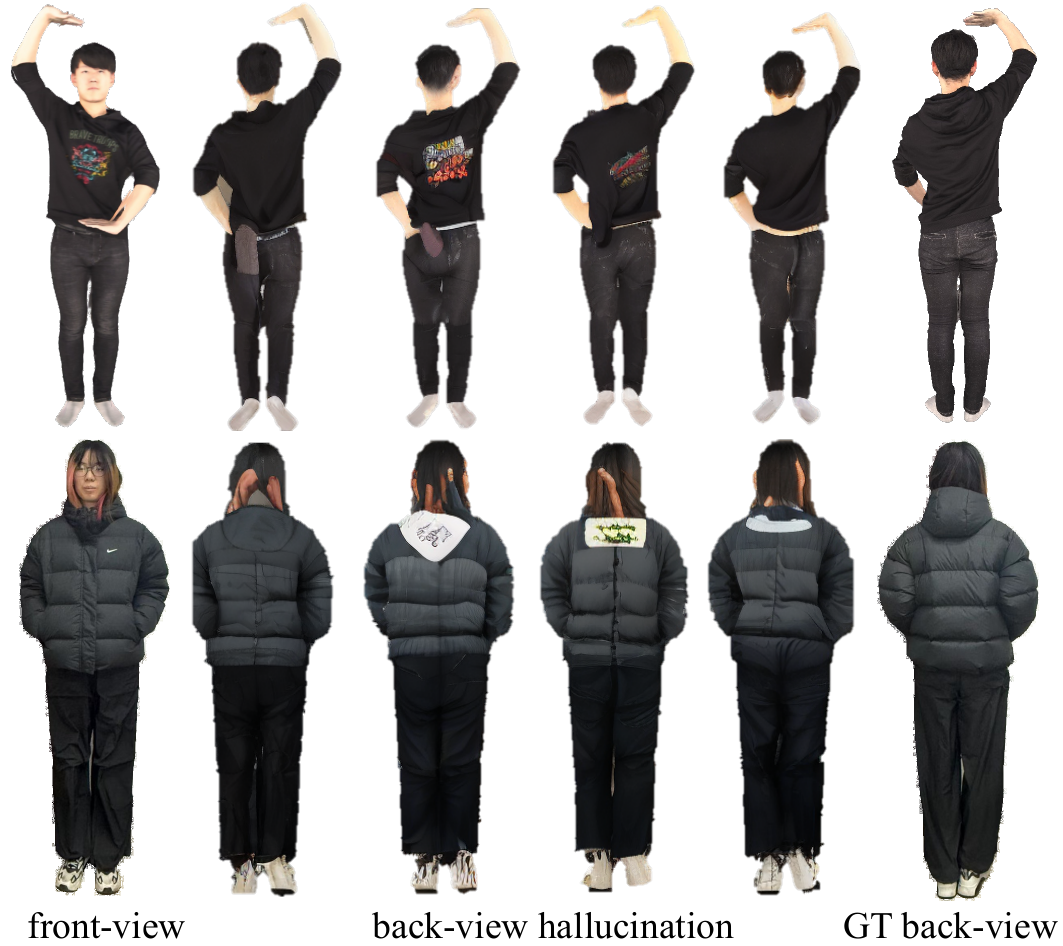}
    \caption{Visualization of different texture introduced by generative methods.}
    \label{fig: gen_model}
\end{figure}

As shown in Fig.~\ref{fig: gen_model}, we additionally visualize the texture uncertainty introduced by generative model~\cite{rombach2022sith_diffusion} in single-view reconstruction method SiTH~\cite{ho2024sith}. Even with the same front view is provided, the model may predict inconsistent back-view textures that significantly deviate from the expected appearance. This may lead to uncontrollable reconstruction results and a noticeable misalignment with the target subject.

\subsection{Ablation Studies on Gaussian Regression}

Regarding Gaussian attribute regression, considering the potential inconsistency between the input front/back views and the side views generated by wrapping based on nearest neighbor search (NNS), we conduct an ablation study on the number of regression network, as shown in Tab.~\ref{tab: regression_head}. The results show that adding the regression network for the side views specially brings no significant improvement to the human reconstruction, which also verifies the effectiveness of ours proposed side-views enhancement module.

\subsection{Foundation Geometry Prior}
To validate the importance of DUSt3R's~\cite{dust3r_cvpr24} geometry priors, we present ablation experiments on loading foundation geometry reconstruction prior in Tab.~\ref{tab: pretraining}. The geometry priors of DUSt3R improve the reconstruction results of our method, which demonstrates the importance of introducing general foundation priors into the specific human domain.

\begin{table}[t]
\centering
\begin{tabular}{c|ccc}
\toprule
 fusion strategy & PSNR $\uparrow$ & SSIM $\uparrow$ & LPIPS $\downarrow$\\
\midrule
 Concat & 22.40 & 88.76 & \textbf{13.29} \\
 Average & \textbf{22.44} & \textbf{88.78} & 13.47 \\
\bottomrule
\end{tabular}
\caption{Comparison on fusion strategy of side-view tokens.}
\label{tab: average_vs_concat}
\end{table}

\subsection{Token Fusion Strategy}

In Sec. 3.3 of the main paper, we obtain global side-view tokens by averaging the front and back image tokens. In Tab.~\ref{tab: average_vs_concat} we present ablation experiments comparing the averaging and concatenation strategies for obtaining side-view tokens. We observe that both methods yield good reconstruction quality, which demonstrates the effectiveness of the point cloud prediction model.

\subsection{Reconstruction results from arbitrary inputs.}

\begin{figure}[t]
    \centering
    \includegraphics[width=\linewidth]{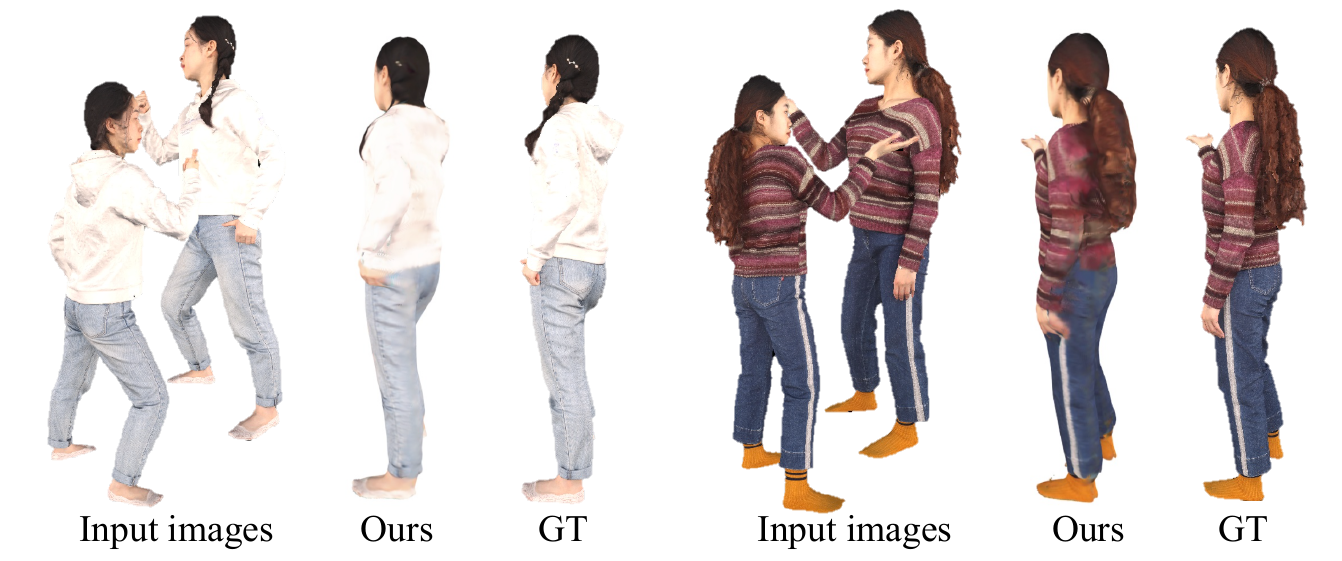}
    \caption{Reconstruction results based on arbitrary image inputs.}
    \label{fig: side_view}
\end{figure}

We provide visualization results of human reconstruction using two input images from two views apart from front and back views in Fig.~\ref{fig: side_view}. As shown in the figure, our method is able to produce high-quality reconstructions even from arbitrary image input, demonstrating strong generalization capability.

\subsection{Multi-view Scenario}

\begin{figure*}[thbp]
    \centering
    \includegraphics[width=1\linewidth]{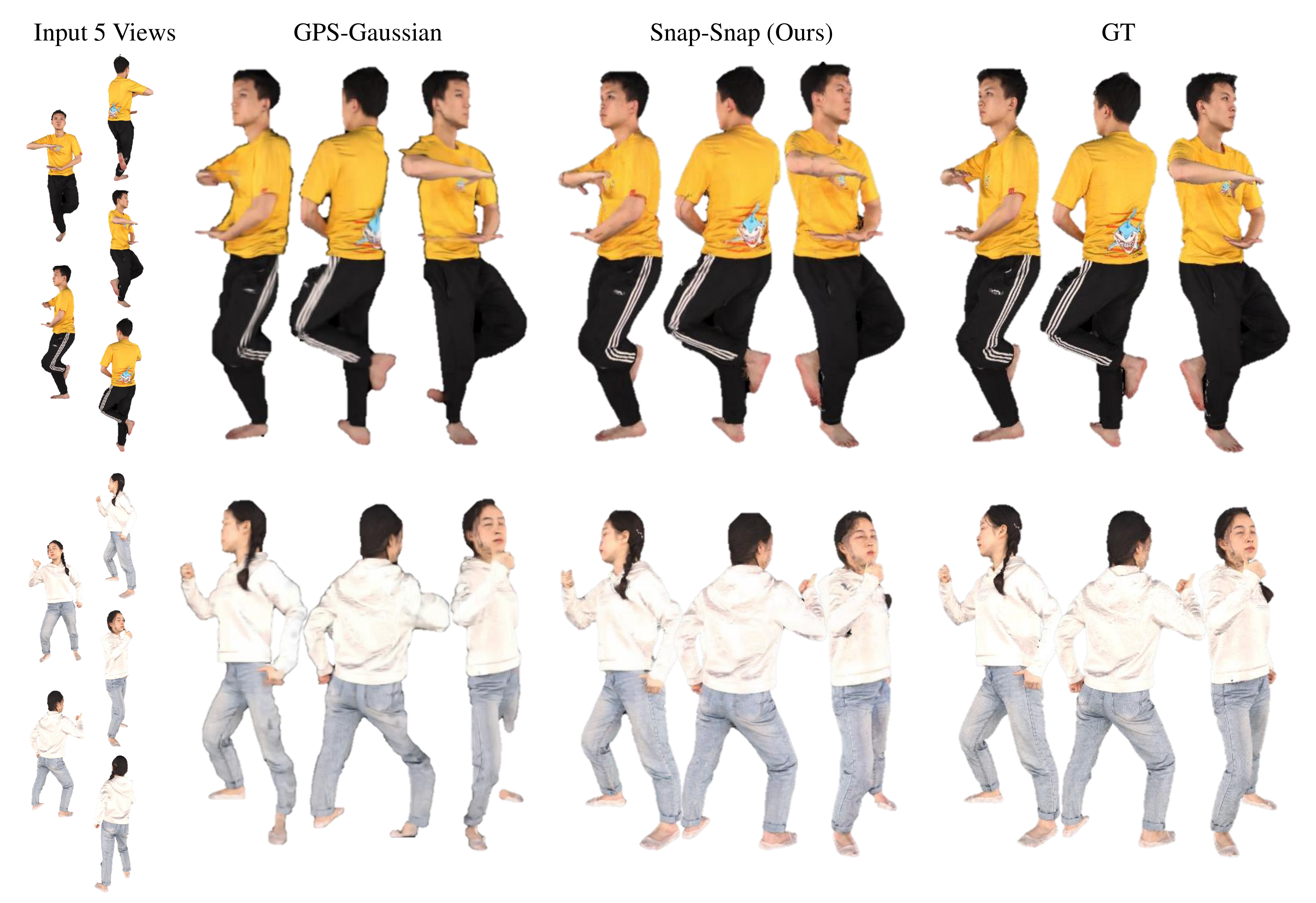}
    \caption{We visualize the comparison of reconstruction results between our method and GPS-Gaussian~\cite{zheng2024gps} with five input views.}
    \label{fig: vis_5views}
\end{figure*}

To further evaluate our reconstruction performance, we conduct experiments in a scenario with five input views and compare it with GPS-Gaussian~\cite{zheng2024gps}, as shown in Tab.~\ref{tab: 5views} and Fig.~\ref{fig: vis_5views}. Considering that it is different from the two-view scenario, there is no information lacking in the scenario with 5 input views. We only predict the point clouds in the two input views, removing the part that predicts the side view point clouds. Our algorithm achieves high-definition reconstruction under multiple views. We conduct quantitative and qualitative comparisons with GPS-Gaussian. It can be observed that our reconstruction method not only achieves complete human body reconstruction in multi-view input scenarios but also attains higher reconstruction quality.

\begin{table}[t]
\centering
\begin{tabular}{c|ccc}
\toprule
Method & PSNR $\uparrow$ & SSIM $\uparrow$ &  \begin{tabular}{c}LPIPS $\downarrow$ \\\end{tabular} \\
\midrule
 GPS-Gaussian~\cite{zheng2024gps} & 20.76 & 88.04 & 14.83 \\
 Snap-Snap (Ours) & \textbf{24.73} & \textbf{91.59} & \textbf{10.67} \\
\bottomrule
\end{tabular}
\caption{We compare the reconstruction results of our method with GPS-Gaussian~\cite{zheng2024gps} in a scenario with five input views.}
\label{tab: 5views}
\end{table}

\subsection{Impact of Overlapping Point Clouds.}
\begin{figure}[t]
    \centering
    \includegraphics[width=1\linewidth]{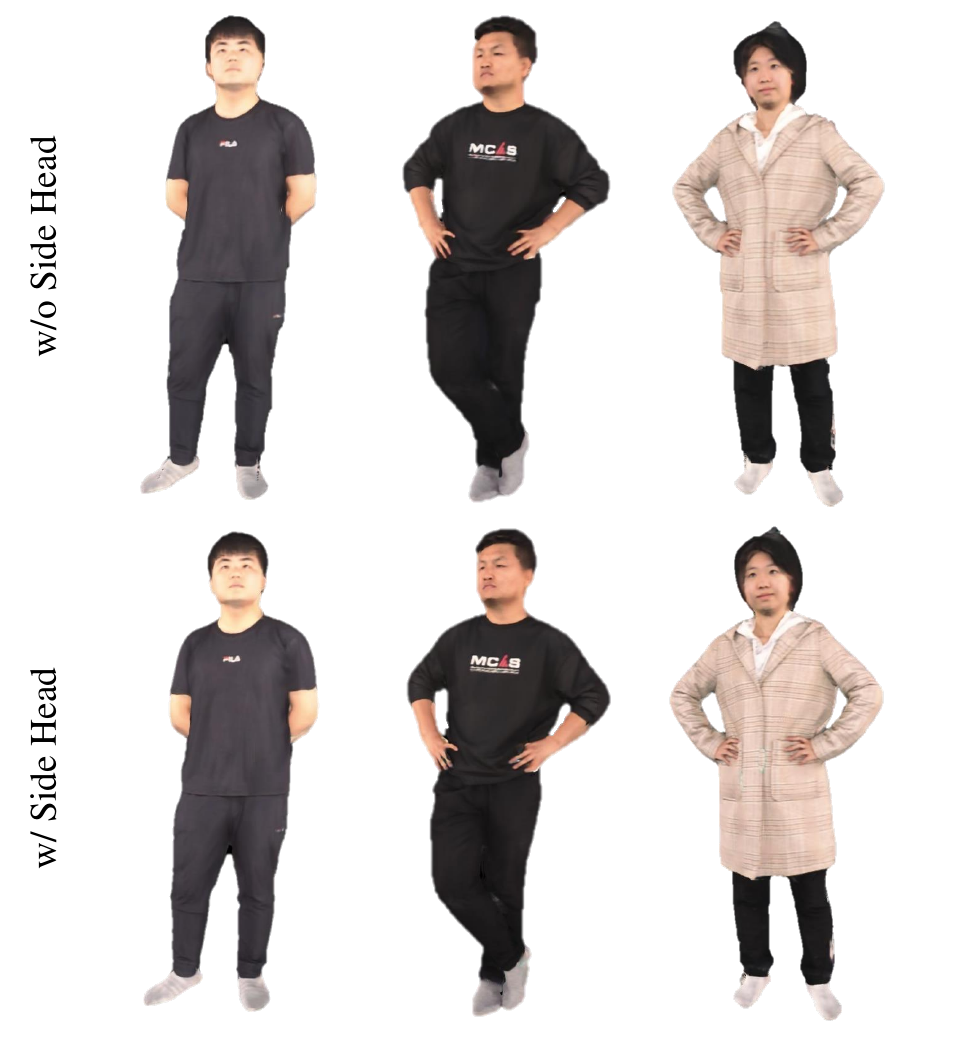}
    \caption{We present the visual results of the impact of the additional side-view heads on the human construction.}
    \label{fig: sup_front}
\end{figure}

\begin{figure}[t]
    \centering
    \includegraphics[width=\linewidth]{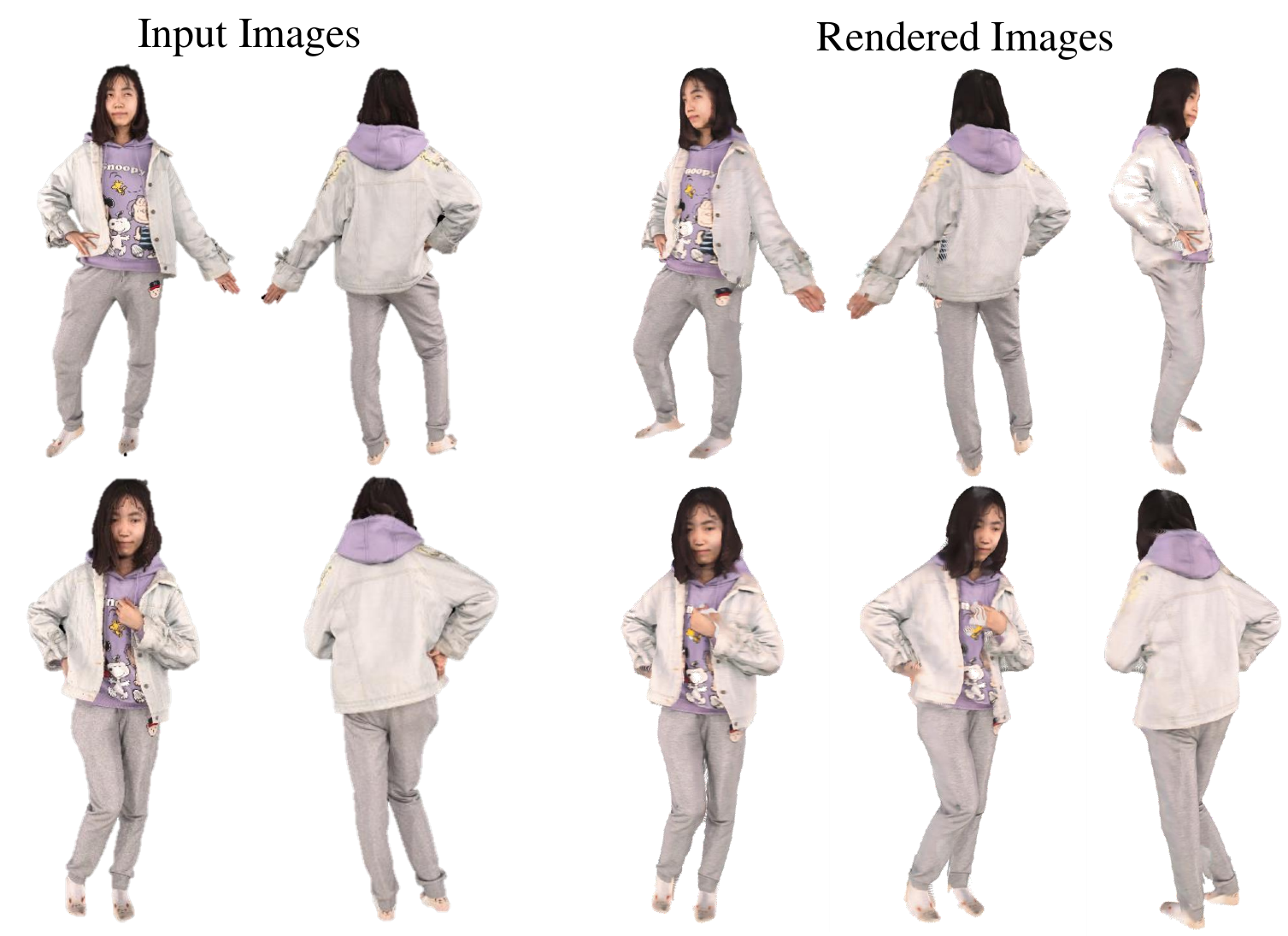}
    \caption{Reconstruction results of the same human in different poses.}
    \label{fig: diff_pose}
\end{figure}

\begin{figure}[t]
    \centering
    \includegraphics[width=1\linewidth]{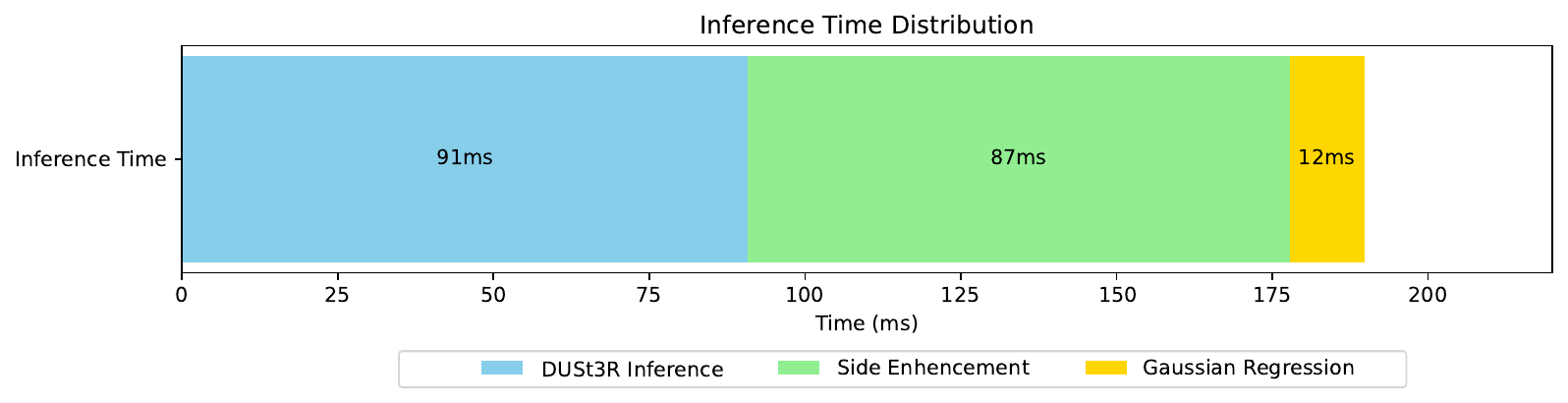}
    \caption{The visual results of the time consumption for different modules in the inference process.}
    \label{fig: infer}
\end{figure}

In the main paper, we adapt additional side-view prediction heads to complete the geometry of the side views. However, it is unavoidable that there is some overlap between the point clouds predicted for the side views and those for the front view, which may affect the quality of front view reconstruction. Therefore, in Fig.~\ref{fig: sup_front}, we present visualization results of front view reconstruction before and after incorporating side-view prediction heads. We can observe that after adding side-view prediction heads, although there are some additional overlapping point clouds in the front view, this has almost no impact on the front view reconstruction quality.

\subsection{Consistency of Reconstruction}
To validate the consistency of our method in reconstructing human bodies with different poses, we present reconstruction results of the same human in various poses in Fig.~\ref{fig: diff_pose}. We can observe that our method achieves good reconstruction quality across different poses while maintaining excellent consistency.

\subsection{Time-consuming Analysis}
In Fig.~\ref{fig: infer} we show the inference time spent by each module of our model, with the total inference time around 190 ms.

\subsection{Network Architecture}

\begin{figure*}[thbp]
    \centering
    \includegraphics[width=0.9\linewidth]{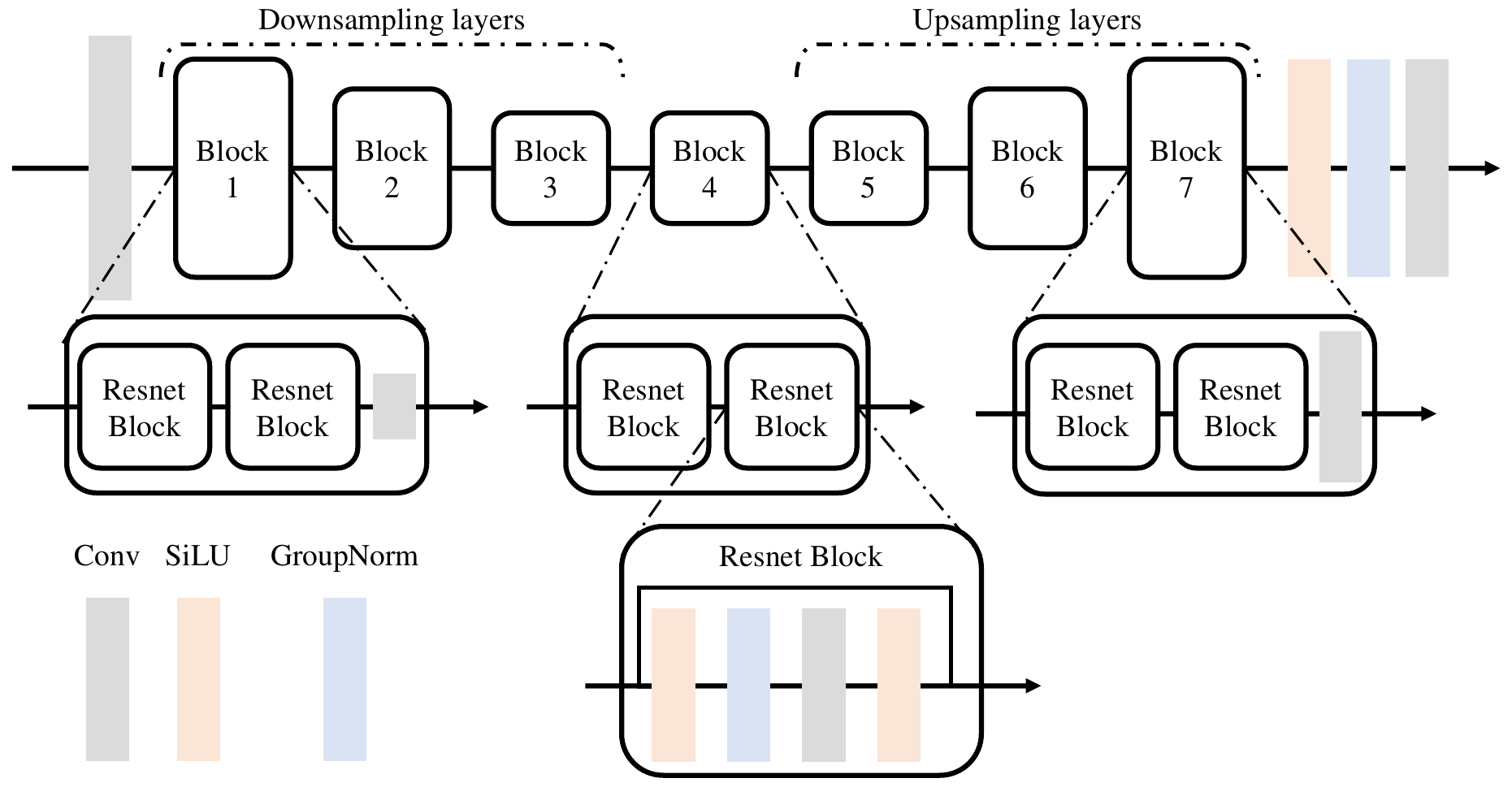}
    \caption{The network architecture of the Gaussian attribute regression network.}
    \label{fig: vis_unet}
\end{figure*}

We use a UNet-like~\cite{ronneberger2015u} network architecture in the Gaussian attribute regression. The specific architecture is shown in the Fig.~\ref{fig: vis_unet}. For the Gaussian attribute regression network, the inputs are the images from four viewpoints, which are concatenated together and fed into the network. The output is the corresponding Gaussian attributes for the corresponding view.
Considering the resolution of input images, we only use ResNet blocks~\cite{he2016deep} to form the UNet-like network. The downsampling layer number and upsampling layer number are 3. For Gaussian attribute regression network, we use convolution dimensions of (16, 16, 16).

\begin{figure*}[thbp]
    \centering
    \includegraphics[width=1\linewidth]{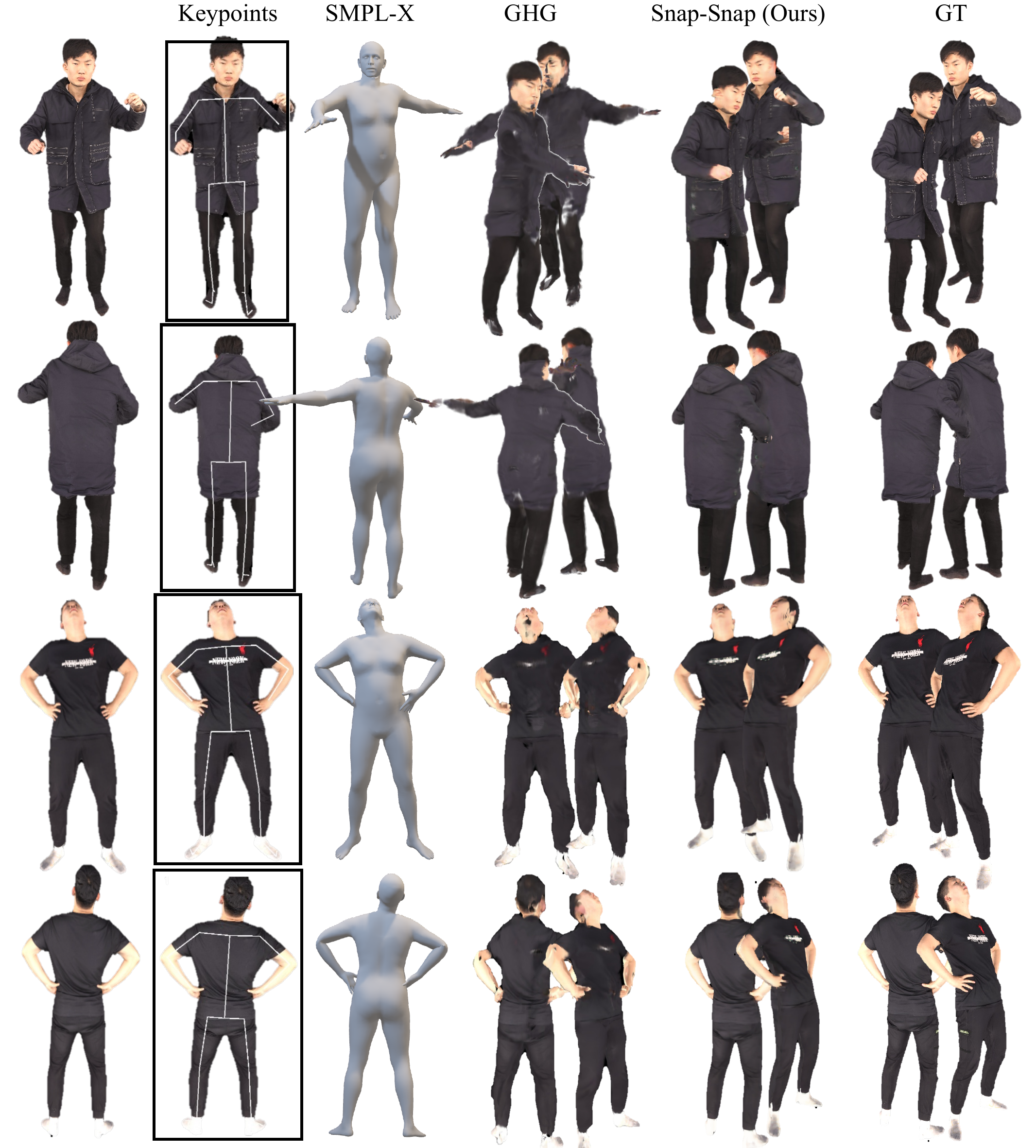}
    \caption{We visualize the reconstruction results of GHG~\cite{ghg_2024} based on SMPL-X~\cite{smplx} parameters predicted by EasyMocap~\cite{easymocap}, as well as our reconstruction results in the same scenario. The first column shows the input views, the second column shows the detected keypoints, and the third column shows the corresponding predicted SMPL-X mesh visualization results. Columns 4-9 show the reconstruction results of GHG, our reconstruction results, and the ground truth, respectively.}
    \label{fig: vis_smpl}
\end{figure*}

\subsection{Discussion on Snap-Snap}
We observe that the reconstruction results of Snap-Snap still contain some holes, particularly around the armpits or regions occluded by the arms. These missing parts in the human Gaussians are due to the limitations of point cloud supervision, which is derived from depth maps~\cite{dust3r_cvpr24}, wherein certain points are filtered out because of occlusions. The hollows on the human bodies could potentially be reduced with the use of geometric generative priors. As the occluded areas are relatively small, they have little effect on the consistency of the final reconstruction.

\subsection{Discussion on GHG~\cite{ghg_2024}}

We further visualize the SMPL-X~\cite{smplx} parameters predicted based on EasyMocap~\cite{easymocap} in Fig.~\ref{fig: vis_smpl}, and it can be observed that estimating human body parameters using only two views is very challenging. The visualization results show that the predicted SMPL-X parameters have a significant impact on reconstruction methods based on the SMPL-X model, such as GHG~\cite{ghg_2024}. In contrast, our method reconstructs the human body based on directly predicted point clouds, effectively avoiding inaccuracies in SMPL-X parameter estimation.

\end{document}

%% file: preamble.tex
\usepackage[dvipsnames]{xcolor}

\usepackage{graphicx}
\usepackage{amsmath}
\usepackage{amssymb}
\usepackage{booktabs}
\input{math}

\usepackage{algorithm}
\usepackage{algpseudocode}
\usepackage{times}
\usepackage{epsfig}
\usepackage{epigraph}
\usepackage{graphicx}
\usepackage{amsmath}
\usepackage{amssymb}
\usepackage{tabularx}
\usepackage{multirow}
\usepackage{float}
\usepackage{soul}
\usepackage{pifont}
\usepackage{wrapfig}
\usepackage{threeparttable}


%% file: math.tex
\usepackage{amsmath,amsfonts,bm}

\def\eqref#1{equation~\ref{#1}}

\def\1{\bm{1}}

\DeclareMathAlphabet{\mathsfit}{\encodingdefault}{\sfdefault}{m}{sl}
\SetMathAlphabet{\mathsfit}{bold}{\encodingdefault}{\sfdefault}{bx}{n}

